\documentclass[conference,compsoc]{IEEEtran}
\usepackage{tikz}
\usepackage{textcomp}

\ifCLASSOPTIONcompsoc
  \usepackage[nocompress]{cite}
\else
  \usepackage{cite}
\fi
\ifCLASSINFOpdf
\else
\fi
\usepackage{amsthm}
\usepackage{amsmath}
\usepackage{amssymb}
\newtheorem{lemma}{Lemma}
\newtheorem{theorem}{Theorem}

\newtheorem*{remark}{Remark} 
\newtheorem{definition}{Definition}

\usepackage{bm}
\usepackage{bbm}

\usepackage{xcolor}
\definecolor{color_orange}{RGB}{247, 149, 63}
\definecolor{color_green}{RGB}{88, 178, 88}
\definecolor{color_gray}{RGB}{230, 228, 235}
\definecolor{color_bg}{RGB}{241, 243, 246} 
\definecolor{color_text}{RGB}{110, 115, 125}
\usepackage{algorithmic}
\newcommand{\algrule}[1]{\par\vspace{2pt}\hrule height #1\vspace{2pt}\par}

\usepackage{booktabs} 
\usepackage{pgfplots}
\usepgfplotslibrary{groupplots}
\pgfplotsset{compat=1.15}

\usepackage{enumitem}
\usepackage{multirow} 

\usepackage[hidelinks]{hyperref}
\usepackage{url}
\IEEEoverridecommandlockouts

\begin{document}
%
% paper title
% Titles are generally capitalized except for words such as a, an, and, as,
% at, but, by, for, in, nor, of, on, or, the, to and up, which are usually
% not capitalized unless they are the first or last word of the title.
% Linebreaks \\ can be used within to get better formatting as desired.
% Do not put math or special symbols in the title.
%\title{vPIR with Identical Online Cost to SimplePIR}
\title{Can Verifiable PIR Match the Online Efficiency of SimplePIR?}

% author names and affiliations
% use a multiple column layout for up to three different
% affiliations
%\author{\IEEEauthorblockN{Xinxuan Zhang}
%\IEEEauthorblockA{State Key Laboratory of Cyberspace Security Defense, Institute of Information Engineering, CAS\\
%\and
%\IEEEauthorblockN{Homer Simpson}
%\IEEEauthorblockA{Twentieth Century Fox\\
%Springfield, USA\\
%Email: homer@thesimpsons.com}
%\and
%\IEEEauthorblockN{James Kirk\\ and Montgomery Scott}
%\IEEEauthorblockA{Starfleet Academy\\
%San Francisco, California 96678-2391\\
%Telephone: (800) 555--1212\\
%Fax: (888) 555--1212}}

% conference papers do not typically use \thanks and this command
% is locked out in conference mode. If really needed, such as for
% the acknowledgment of grants, issue a \IEEEoverridecommandlockouts
% after \documentclass

% for over three affiliations, or if they all won't fit within the width
% of the page (and note that there is less available width in this regard for
% compsoc conferences compared to traditional conferences), use this
% alternative format:
% 
\author{\IEEEauthorblockN{Xinxuan Zhang\IEEEauthorrefmark{2}\IEEEauthorrefmark{3},
Ruida Wang\IEEEauthorrefmark{1}\IEEEauthorrefmark{2}\IEEEauthorrefmark{3}\thanks{\IEEEauthorrefmark{1}Corresponding author.},
Zirui Wang\IEEEauthorrefmark{2}\IEEEauthorrefmark{3}, 
Yuming Guo\IEEEauthorrefmark{2}\IEEEauthorrefmark{3},
Jikang Bai\IEEEauthorrefmark{2}\IEEEauthorrefmark{3},\\
Yuanju Wei\IEEEauthorrefmark{2}\IEEEauthorrefmark{3},
Yi Deng\IEEEauthorrefmark{4} and
Xianhui Lu\IEEEauthorrefmark{2}\IEEEauthorrefmark{3}}
\IEEEauthorblockA{\IEEEauthorrefmark{2}State Key Laboratory of Cyberspace Security Defense, \\Institute of Information Engineering, CAS, Beijing, China}
\IEEEauthorblockA{\IEEEauthorrefmark{3}School of Cybersecurity, UCAS, Beijing, China}
\IEEEauthorblockA{\IEEEauthorrefmark{4}Xidian University, Xi\textquotesingle an, China}
\IEEEauthorblockA{Email:\IEEEauthorrefmark{2}\{zhangxinxuan, wangruida, wangzirui, guoyuming, baijikang, weiyuanju, luxianhui\}@iie.ac.cn\\
\IEEEauthorrefmark{4}ydeng.cas@gmail.com }}

% use for special paper notices
%\IEEEspecialpapernotice{(Invited Paper)}

% make the title area
\maketitle

% \IEEEcopyrightnotice
% As a general rule, do not put math, special symbols or citations
% in the abstract
\begin{abstract}

Verifiable Private Information Retrieval (vPIR) ensures that a malicious server cannot return tampered database records, but it traditionally incurs significant overhead compared to classical PIR. Recent advancements, notably VeriSimplePIR, approach the online efficiency of SimplePIR by introducing a proof preprocessing phase and requiring only lightweight client-side online verification. However, this scheme relies on an additional Short Integer Solution (SIS) assumption and SIS-based commitments, which only support \emph{relaxed} range proofs. Consequently, these schemes are forced to adopt significantly weaker parameter choices, resulting in additional online overhead.

In this paper, we propose a novel vPIR protocol that employs a newly designed error-correcting code-based commitment scheme as the database digest. This approach (i) supports efficient proofs of linear relations, (ii) achieves \emph{precise} range proofs and (iii) eliminates the reliance on the SIS assumption. Thus, our scheme successfully adopts parameter choices identical to those of SimplePIR. 

As a result, our vPIR protocol is the first to achieve exactly the same online communication and server computation costs as SimplePIR, while keeping online client computation nearly identical. Additionally, compared to VeriSimplePIR, our protocol reduces online communication and client-side computation by approximately 20\%, and decreases online server-side computation by over 35\%.
\end{abstract}

\IEEEpeerreviewmaketitle

\section{Introduction}

Private Information Retrieval (PIR) addresses a fundamental privacy problem in outsourced data storage: how can a client retrieve an element from a remote database without revealing to the server which element was requested? Informally, a PIR protocol allows a client to obtain the $i$-th entry of an $n$-item database held by one (or more non-colluding) server(s), while ensuring that the server(s) learn no information about the index $i$. This capability is essential for applications where query privacy is critical, including anonymous web browsing, privacy-preserving patent or medical record searches, location-based services, and secure cloud storage. By decoupling data access from access pattern leakage, PIR provides a rigorous cryptographic foundation for protecting user privacy in the face of untrusted service providers.

However, standard PIR protocols assume that the server executes the retrieval protocol honestly (though it might be curious), returning the correct database entry for the requested index. In practice, a malicious or faulty server may deviate from the protocol — returning arbitrary, outdated, or corrupted data — without the client being able to detect the misbehavior. This vulnerability is particularly critical in settings where the server has an incentive to manipulate query results, such as adversarial cloud providers or compromised nodes in a distributed system. To address this issue, verifiable Private Information Retrieval (vPIR) extends the classical PIR model by adding an integrity guarantee: the client can efficiently verify that the retrieved value is indeed the correct entry corresponding to its query, even if the server behaves maliciously. vPIR preserves the privacy requirement (the server learns nothing about the client’s index) while simultaneously providing public or client-side verifiability of the result’s correctness. %In this paper, we focus on VPIR in the single‑server setting as the primary object of study.%, where the server provides a database digest and offers integrity guarantees that the returned response is additionally consistent with that digest.

%Early VPIR schemes \cite{WangZ18,ZhangS14} did not consider selective failures, where a malicious server attempts to learn  the client's query based on whether the client accepts the proof. These schemes only provided authenticity guarantees that a response is consistent with some valid database, rather than with the specific database to which the server committed a priori. Colombo et al.~\cite{CNCWF23} addressed the issue of selective failures. Their Authenticated PIR scheme is the first VPIR construction that resists selective‑failure attacks. In this scheme, the server sends an honest digest of the database before the protocol begins and, when responding, ``proves'' the consistency of the response with the digest.

%Later VPIR schemes\cite{DietzT24,CastroL24,RLP26} guarantee security even if the server maliciously generates the digest. %The two most efficient among them are VeriSimplePIR\cite{CastroL24} and Rathee et al.'s scheme\cite{RLP26}. VeriSimplePIR, derived from SimplePIR, is one of the most online‑efficient VPIR schemes to date. Very recently, Rathee et al.'s scheme, based on Hintless, eliminates the need for large client‑side storage of SimplePIR, at a cost of larger online overhead.

The primary challenge in vPIR design is maintaining practical performance, particularly during the online phase, as this directly dictates the query response latency experienced by the client. Historically, achieving verifiability has incurred prohibitive computational and communication overhead compared to baseline semi-honest PIR protocols. Recent advancements, notably VeriSimplePIR \cite{CastroL24}, attempt to address this bottleneck by introducing a proof preprocessing phase and requiring only lightweight client-side verification. This approach aims to mirror the online efficiency of SimplePIR~\cite{10.5555/3620237.3620455} --- a highly server-efficient semi-honest PIR scheme that set a critical benchmark for online performance by offloading heavy cryptographic workloads to an offline preprocessing phase.

However, upgrading such semi-honest PIR protocols to resist malicious servers first requires cryptographically binding a public database digest to the server's online homomorphic computations. To achieve this binding, VeriSimplePIR heavily relies on the Short Integer Solution (SIS)\cite{10.1145/237814.237838} assumption and SIS-based commitments.  This architectural choice introduces a fundamental mathematical bottleneck: SIS-based commitments inherently support only \emph{relaxed}, rather than \emph{precise}, range proofs. Consequently, the protocol can only guarantee a loose upper bound on the database entries. Since homomorphic evaluation over the underlying Regev encryption scheme amplifies ciphertext noise proportionally to the norm of these entries, this loose bound forces the system to account for an inflated noise growth. To ensure decryption correctness, the scheme is forced to adopt significantly weaker parameter choices. This parameter compromise directly translates into substantial online overhead, specifically resulting in a $\approx$40\% slowdown in server computation time and a 40\% to 50\% increase in communication costs compared to the SimplePIR baseline. 

Other contemporary schemes attempt to optimize different dimensions of vPIR but further exacerbate the online efficiency bottleneck. Very recently, Rathee et al.'s scheme \cite{RLP26} builds upon hintless PIR techniques to eliminate the large client-side storage requirements inherent to VeriSimplePIR. While successful in reducing client storage, this approach incurs a penalty during online execution, resulting in a 2.5$\times$ cost in server-side cost than VeriSimplePIR~\cite{CastroL24}.

Since all existing vPIR schemes incur inherent online overhead—either directly to support verification mechanisms or indirectly to offset parameter compromises—a critical question emerges: to what extent can this online penalty be minimized? More specifically:\medskip

\textit{Can verifiable PIR match the online efficiency of SimplePIR?}

%Both schemes described above are based on lattice assumptions. They employ Regev's LWE cryptosystem to achieve privacy, and the server uses an SIS hash commitment as the digest of the database. However, this commitment suffers from two drawbacks:
%\begin{enumerate}[leftmargin=*]
%    \item The commitment and its accompanying proof system cannot precisely enforce the range of $D$. Specifically, for a bound $D < \beta$, the commitment and proof can only guarantee that the committed value $D'$ satisfies $D' < k\beta$ for a large $k$. This limitation forces the parameter selection of Regev's LWE cryptosystem to adopt looser parameters to ensure successful decryption.
%    \item 
    %The commitment itself is excessively large. For a 16~GB database, the commitment size reaches as high as 1~GB. Although Rathee et al.'s scheme improves this such that the client no longer needs to store the commitment during the online phase, it still requires the client to download the full commitment and perform the associated verification during the preprocessing phase. This incurs prohibitively high communication and computation overheads in the preprocessing stage, making the scheme highly unfriendly to clients with a limited number of queries or constrained computational/storage resources.
%\end{enumerate}

\subsection{Our Result.}

%We construct a new VPIR scheme that achieves lower communication overhead in the offline phase and higher efficiency in the online phase. Our scheme adopts a novel proof paradigm based on tensor IOP and a commitment mechanism via encoding. This approach eliminates the need to send the SIS hash commitment $H$ while still enabling verification of the response's correctness. Furthermore, we provide a more precise proof algorithm for bounding the size of $D$, ensuring that the online phase can employ the same encryption parameters as those used in the semi‑honest setting.

In this paper, we answer this affirmatively by proposing a novel vPIR protocol through a fundamental theoretical advancement: constructing the first practical vPIR protocol whose malicious security is grounded exclusively on the standard Learning with Errors (LWE) assumption—bringing the security foundation of verifiable PIR back in line with most modern PIR schemes.\medskip

\noindent \textbf{Theoretical Result.} State-of-the-art verifiable PIR schemes (notably VeriSimplePIR~\cite{CastroL24} and Rathee et al.'s scheme \cite{RLP26}) fundamentally rely on a hybrid of LWE and SIS assumption. They employ LWE-based encryption to ensure privacy-preserving computation, and SIS-based commitments to achieve verifiability. However, this paradigm inherently restricts them to relaxed data range proofs and forces a significant degradation in cryptographic parameter choices. We propose a novel vPIR architecture {decouples verifiability from the SIS assumption, thereby eliminating the need for it} and theoretically streamlining the security foundation. 

Crucially, this theoretical simplification directly translates into practical gains. Unburdened by SIS-induced parameter bloat, our scheme seamlessly unlocks and adopts the optimal cryptographic parameters of the semi-honest SimplePIR baseline.

\medskip

\noindent \textbf{Performance Highlight.} 
As a result, our vPIR protocol is the first to achieve exactly the same online communication and server computation costs as the semi-honest SimplePIR, while keeping the online client computation virtually identical.  In concrete terms, our implementation delivers an end-to-end online query latency of merely 69 ms, 115 ms, 358 ms, 657 ms, and 1244 ms for database sizes scaling across 512 MB, 1 GB, 4 GB, 8 GB, and 16 GB, respectively—establishing a new benchmark for online efficiency among verifiable PIR schemes.

Compared to VeriSimplePIR, our protocol reduces online communication and client-side computation by approximately 20\%, and decreases online server-side computation by over 35\%. This performance gap widens even further when compared to the very recent Rathee et al. scheme, which suffers a 2.5$\times$ online cost penalty (relative to VeriSimplePIR) in exchange for reduced client storage. 

%In this paper, we propose a novel vPIR protocol that employs a newly designed error-correcting code-based commitment scheme as the database digest. This approach i) supports efficient proofs of linear relations, ii) achieves \emph{precise} range proofs and iii) eliminates the reliance on the SIS assumption. Thereby, our scheme successfully adopts parameter choices identical to those of SimplePIR. 

%As a result, our vPIR protocol first achieves exactly the same online communication and server computation costs as SimplePIR, while keeping online client computation nearly identical. Additionally, compared to VeriSimplePIR, our protocol reduces online communication and client-side computation by approximately 20\%, and decreases online server-side computation by over 35\%.

\subsection{Related Work}

Since the seminal work of Chor et al.~\cite{492461}, PIR has evolved significantly, though historically plagued by prohibitive communication and computation costs. Modern advancements have largely bifurcated into optimizing semi-honest efficiency and addressing the rigorous demands of malicious security. \medskip

\noindent \textbf{Semi-Honest PIR. } Recent years have seen a surge in highly practical PIR schemes operating under the semi-honest setting ~\cite{9833700,10.1007/978-3-031-68400-5_6,10.1007/978-3-031-91095-1_1,3698900.3699235,10.1145/3658644.3690328,10.1145/3564246.3585175,10.1007/978-3-030-45721-1_3}. A prominent online efficiency benchmark in this domain is SimplePIR \cite{10.5555/3620237.3620455}. It offloads the computationally heavy homomorphic computations to an one-time setup phase. By doing so, it achieves exceptionally low online communication and high server throughput. Nevertheless, SimplePIR implicitly trusts the server to execute operations honestly, leaving the client entirely vulnerable to malicious manipulation or corrupted data delivery.\medskip

\noindent \textbf{Early Efforts in Verifiable PIR. }The challenge of securing PIR against malicious servers was initially conceptualized by Kushilevitz and Ostrovsky~\cite{646125}. Subsequently, various cryptographic approaches have explored this domain. However, early vPIR protocols ~\cite{10.1007/978-3-031-22368-6_1,ZHAO2021897} primarily focused on ensuring that the server's response was honestly generated from \emph{some} database, fundamentally failing to prevent the server from swapping the database dynamically during the query phase. Consequently, these schemes exhibit low resistance to \emph{selective failure attacks} and lack the cryptographic binding necessary for modern outsourced database integrity. To cryptographically bind the server to a specific database, modern vPIR constructions \cite{CastroL24,RLP26} require a robust database digest setup, by introducing an extractable SIS commitment as discussed above.

\section{Technical Overview}
\noindent\textbf{Background: SimplePIR and VeriSimplePIR.} SimplePIR presents a very simple PIR construction based on Regev encryption. For a database \(D \in \mathbb{Z}_p^{\ell \times m}\) (we write each entry of \(D\) in the range \((-p/2, p/2]\) so that \(\|D\|_{\infty} \le p/2\)), and a query vector \(\mu\) (i.e., the client wishes to retrieve \(D\mu\)), the client computes a Regev encryption $(A,u)$ of \(\mu\) and sends it to the server. The server then computes and returns \((H=DA,v = Du)\). The client decrypts to obtain \(D\mu\). To reduce online overhead, the scheme precomputes \(H = DA\) during the setup phase, as the choice of \(A\) is independent of the query, and it can be reused for all queries. In the online phase, the client only needs to compute \(u\), the server returns \(v = Du\), and the client decrypts using \((H, v)\) to obtain \(D\mu\).

%At the begining, all parties use a public random matrix \(A \gets \mathbb{Z}_q^{m \times n}\) as the masking matrix for Regev encryption. In the setup phase, for a database \(D \in \mathbb{Z}_p^{\ell \times m}\), the server outputs \(H = DA\) as a hint. In the query phase, when the client wishes to retrieve \(D\mu\), the client computes and sends an encryption \(u = As + e + \textsf{Ecd}_{p,q}(\mu)\) using the Regev encryption scheme. The server then computes and returns \(v = Du\). Finally, the client can decrypt \((H s, v)\) to obtain \(D\mu\).

%To achieve verifiability for SimplePIR, the server needs to set a digest of $D$ in the setup phase, and then two core guarantees are required: (1) the digest fixes a unique \(D\), and \(H = DA\), \(v = Du\); (2) the entries of \(D\) are sufficiently small to ensure the success of Regev decryption in the final step. 

To achieve verifiability for SimplePIR, the server must first generate a digest of \(D\) during the setup phase. Two core guarantees are then required: 
\begin{enumerate}
    \item the digest binds to a unique matrix \(D\) that satisfies \(H = DA\) and \(v = Du\); 
    \item the entries of \(D\) are sufficiently small to ensure the success of Regev decryption in the final step.
\end{enumerate}

%VeriSimplePIR scheme use a (extractable) SIS commitments to achieve above requirements. Specifically, by treating \(A\) as the public parameter matrix in the SIS assumption at the same time, \(H\) becomes an SIS commitment binding \(D\) (two small-norm matrices \(D\) and \(D'\) satisfying \(H = DA = D'A\) yields a solution \((D - D')A = 0\) to the SIS problem). Then the client sends an encrypted reusable random challenge matrix \(C\gets \{0,1\}^{\lambda\times \ell}\) to the server, who homomorphically responds with  \(Z = CD\). Then, to ensure that \(H = DA\) and \(v = Du\), the client only need to checks that \(\|Z\|_\infty \le \ell p\) and that \(CH = ZA\). To ensure the small-norm of $D$, the client only checks that \(\|Z\|_\infty \le \ell p\)

VeriSimplePIR uses an extractable Short Integer Solution (SIS) commitment to achieve the above requirements. Specifically, by treating \(A\) as the public parameter matrix in the SIS assumption, \(H\) becomes an SIS commitment that binds \(D\): if two small‑norm matrices \(D\) and \(D'\) satisfy \(H = DA = D'A\), then \((D - D')A = 0\) yields a solution to the SIS problem.
The client then sends a \textit{encrypted} random challenge matrix \(C \gets \{0,1\}^{\lambda \times \ell}\) (which can be reused) to the server, who homomorphically responds with \(Z = CD\)\footnote{To ensure the correctness of \(Z = CD\), the original VeriSimplePIR scheme additionally used an SIS commitment to \(D^\perp\). Subsequent work by Rathee et al. pointed out that the SIS assumption together with \(CH = ZA\) can already guarantee \(Z = CD\). However, this further affects parameter selection and reduces the efficiency of the online phase.}. Then, to verify that \(H = DA\) and \(v = Du\), the client only needs to check that  \(C [H | v] = Z[A| u]\). To ensure the small-norm of $D$, the client only checks that \(\|Z\|_\infty \le \ell p/2\), which ensures that $\|D\|_{\infty}\le \ell p$.

Most of the protocol's operations are independent of the query and reusable; hence, they can be moved to the setup phase or the proof preprocessing phase. The online phase remains almost identical to that of SimplePIR, with the only difference being that the client additionally needs to check \(Cv = Du\).

%Let the client sends C to the server, who then response $Z=CD$. The client checks $CH=ZA$ and $\|Z\|_\infty\le \ell p$, which ensures that the committed $\|D\|_{\infty}$.

%If two small-norm matrices \(D\) and \(D'\) satisfy \(H = DA = D'A\), then \((D - D')A = 0\) yields a solution to the SIS problem. To ensure that \(H = DA\) and \(v = Du\) refer to the same \(D\), the client sends a random matrix \(C\), and the server returns \(Z = CD\). The client then verifies that \(C[H \mid v] = Z[A \mid u]\).

However, VeriSimplePIR relies on the SIS-based commitments and only achieves a \emph{relaxed} range proof. Consequently, they are forced to adopt significantly weaker parameter choices, resulting in additional online overhead. More specifically, the SIS assumption additionally introduces the following two constraints:
\begin{gather*}
    q\ge 2\ell p\sqrt{m} \\
    2^{2\sqrt{n\log q\log\delta_{sis}}}\ge 
    2\ell p\sqrt{m}
\end{gather*}
And the relaxed range-proof and Regev decryption introducing the following constraint: 
$$q\ge 2p^2\cdot \sigma\cdot \ell \cdot \sqrt{2m\ln(2/\delta)},$$
where $q$ and $p$ are the Regev ciphertext modulus and plaintext modulus respectively, $A\in \mathbb{F}_{q}^{m\times n}$, $\delta_{sis},\delta,\sigma$ are parameters that are set to $1.005$, $2^{-40}$, and $6.4$ respectively.\medskip

\noindent\textbf{Our Approach.} Instead of using \( H = DA \) as a commitment, which relies on SIS assumptions, we want to adopt a simpler method to generate a commitment to \( D \) and use it as the digest of \( D \), for example by directly employing a Merkle hash. However, this approach hinders the proof of subsequent linear relations. 

To address this, rather than hashing \( D \) directly, we first view \( D \) as a matrix, apply error-correcting code encoding row by row, and finally hash the resulting encoded matrix; the resulting hash values constitute a commitment of $D$. Since now the hash actually constitutes a polynomial commitment to the multilinear extension polynomial of \( D \), we can use the tensor-IOP scheme from Ligero/Brakedown\cite{AHIV17,GLSTW23} to prove the ``well-formedness'' of the commitment and the linear relations among the rows of $D$.

%The row linear relation proof provides us with an efficient way to prove \(H = DA\) and \(v = Du\): by committing to \(D^\perp\), we can directly prove the row linear relation \([H \mid v]^\perp = [A \mid u]^\perp \cdot D^\perp\). However, this approach introduces excessive computational overhead when considering proof proprocessing. Therefore, we still adopt the verification method similar to VeriSimplePIR: the client sends an encrypted challenge matrix \(C\) to the server, who then returns a ciphertext of \(Z = CD\). Here, we use the aforementioned row linear relation proof for \(Z = CD\), noting that this proof can be executed even when the server does not know \(C\) or \(Z\), and avoid expensive additional costs or parameter constraints.

The row linear relation proof offers an efficient way to prove \(H = DA\) and \(v = Du\): by committing to \(D^\perp\), we can directly prove \([H| v]^\perp = [A | u]^\perp \cdot D^\perp\). However, this approach incurs excessive overhead when considering proof preprocessing. Hence, we adopt a method similar to VeriSimplePIR: the client sends an encrypted challenge matrix \(C\) to the server, who returns a ciphertext of \(Z = CD\). We then use the row linear relation proof for \(Z = CD\), noting that this proof can be executed even when the server does not know \(C\) or \(Z\) and avoiding significant extra costs or additional parameter constraints. Now, checking \(C[H | v] = Z[A | u]\) implies that \(H = DA\) and \(v = Du\) for the committed database \(D\), without relying on the SIS assumption any longer.

So far, we still need to prove that the norm of \(D\) is sufficiently small. Although the previous approach—checking \(\|Z\|_\infty \le \ell p/2\)—still guarantees \(\|D\|_\infty \le \ell p\), this bound is too loose. To address this, we propose two new range proof techniques:\medskip

\noindent \textit{First Range-Proof Approach.} We observe that to guarantee successful decryption, it is not necessary to prove an upper bound $p/2$ on the infinity norm of \(D\); an upper bound $p\sqrt{m}/2$ on the Euclidean norm of each row of \(D\) also suffices. Our first proposal is therefore an almost precise range proof for the Euclidean norm of the rows of \(D\).

This method follows from the famous Johnson-Lindenstrauss Lemma~\cite{GHL22,BS22}: {any set of points in a high-dimensional space can be projected into a much lower-dimensional space while approximately preserving all pairwise Euclidean distances within a \textit{constant} factor.}

We can view the database \(D \in \mathbb{Z}^{\ell \times m}\) as a set of \(\ell\) points in an \(m\)-dimensional space, where each point corresponds to a row \(D[i,\cdot]\) of \(D\). To prove the upper bound of \(\|D[i,\cdot]\|_2\), the verifier picks a random projection matrix \(\Pi \gets \mathcal{D}^{m \times 256}\) and sends it to the prover. The prover computes \(D^* = D \cdot \Pi\), mapping each row \(D[i,\cdot]\) from the \(m\)-dimensional space into a low-dimensional space of dimension \(256\). For the mapped rows \(D^*[i,\cdot]\), their Euclidean norms are preserved up to a constant factor. Therefore, the verifier can ensure \(\|D[i,\cdot]\|_2 \le c \cdot p\sqrt{m}/2\) for some \textit{constant} \(c\) by checking the Euclidean norm of \(D^*[i,\cdot]\) is close to $p\sqrt{m}/2$ (the correctness of \(D^* = D \cdot \Pi\) can be verified similarly to \(H = DA\)). In our scheme, \(c = \sqrt{\frac{337}{30}}\). Using this range check, the parameter constraint imposed by the decryption failure rate of Regev becomes
\[
q \ge p^2 \cdot \sigma \cdot \sqrt{\frac{337m \ln(2/\delta)}{15}}.
\]
For the same choice of \(q\), compared to SimplePIR, our choice of \(p\) loses at most 1 bit (and remains identical for some parameters), whereas VeriSimplePIR incurs a loss of up to \(0.5\log \ell\) bits (7–9 bits).\medskip

%Specifically, to prove that $\|D[i,\cdot]\|_2 \le p\sqrt{m}$ for each row $D[i,\cdot]$ of $D \in \mathbb{Z}^{\ell \times m}$, the verifier picks a random specific matrix $\Pi \gets \mathcal{D}^{m \times 256}$ and sends it to the prover, where $\mathcal{D}$ is the distribution that $\Pr[0\gets \mathcal{D}]=0.5$ and $\Pr[-1\gets \mathcal{D}]=\Pr[1\gets \mathcal{D}]=0.25$ . The prover computes and sends $D^* = D\cdot \Pi \mod{q}$ to the verifier\footnote{For an element $w \in \mathbb{F}_q$, we always take its representative in the interval $[-q/2, q/2]$.}, who checks that $\|D^*[i,\cdot]\|_2 \le b\sqrt{337m}$ for each $i$. The verifier then randomly selects and sends a random matrix $C \in \{0,1\}^{\lambda \times \ell}$, and the prover responds $Z=CD$. Both party using the row linear relation proof protocol to verify that $Z=CD \mod q$ is computed correctly, and the verifier checks that $CD^*=Z\cdot \Pi \mod q$, which means that $D^* = D\cdot \Pi \mod{q}$ is computed correctly. 

\noindent\textit{Second Range-Proof Approach.} Since our commitment to \(D\) is also a polynomial commitment to the multilinear extension polynomial $f_{D}$ of \(D\), this allows us to use techniques common in SNARKs, such as lookup arguments, to achieve precise range proofs.

To prove that \(\|D\|_\infty \le p/2\), we list all values in \([p]\) as a table \(\mathbf{t}\). We then use a multilinear-polynomial variant of the cq lookup arguments~\cite{EFG22} to prove that every entry in \(\{f_D(x,y)\}_{(x,y)\in\{0,1\}^{\log\ell + \log m}}\) belongs to the table \(\mathbf{t}\), with minor adjustments to fit our setting. This lookup argument ultimately reduces our range check to verifying a single query of \(f_D(r_x,r_y)=t\) for some $r_x,r_y$, which can be proved using the linear relation proofs of the commitment to $D$.

Using this approach, we can precisely prove that \(\|D\|_{\infty} \le p/2\), thereby achieving the same parameter constraint imposed by the decryption failure rate of Regev as in SimplePIR:
\[
q \ge p^2 \cdot \sigma \cdot \sqrt{m \ln(2/\delta)}.
\]
%parameter selection identical to that of SimplePIR. 
However, compared to the first method, this approach incurs higher computational overhead. Nevertheless, the range check can be performed during the setup phase and thus does not affect the online overhead.

\section{Preliminaries}
%LWE, Regev, PIR, VPIR, simplePIR, and so on.

\noindent\textbf{Notation.} All logarithms $\log(\cdot)$ are to the base 2. A uniform sample $x$ drawn from a finite set $S$ is denoted as $x  \xleftarrow{\$}  S$, while a sample drawn from a probability distribution $\chi$ is denoted as $x \leftarrow \chi$. %The distribution induced by a random variable $X$ is denoted by $\{X\}$. 
We use $[\ell]$ to denote the set $\{0,1, \cdots, \ell-1\}$.  For a vector \(v\), its \(i\)-th component is denoted by \(v[i]\). For a matrix \(D\), its \(i\)-th row is denoted by \(D[i,\cdot]\), its \(j\)-th column by \(D[\cdot,j]\), and its entry at position \((i,j)\) by \(D[i,j]\). The infinity norm of a vector or a matrix is denoted by $\| \cdot \|_\infty$, and the Euclidean norm of a vector is denoted by $\| \cdot \|_2$. We use $\lfloor \cdot \rfloor$ to denote the floor function. For $x \in \mathbb{R}$, the rounding function, we use $\lfloor x \rceil$ to denote $\lfloor x + 1/2 \rfloor$. For $a \in \mathbb{Z}$, the value $a - \lfloor a/q \rceil \cdot q$ is denoted as $[a]_q$. Since we also operate on the components of vectors, all these notations can be naturally extended to vectors element-wise. %We denote a ciphertext encrypting a message $\mu$ as $\mathsf{ct}(\mu)$. This notation applies equally when encrypting a set of messages. We use $\mathsf{poly}(n)$ to denote a polynomial function in $n$, and $\mathsf{negl}(n)$ to denote a negligible function in $n$. 
We use $\kappa$ to denote the statistical security parameter, and $\lambda$ to denote the computational security parameter. %Finally, we use $\equiv_\lambda$ to denote computational indistinguishability with respect to $\lambda$.

\subsection{Regev’s LWE Cryptosystem}
%vPIR  incorporates the core ideas of Regev's LWE cryptosystem, which we introduce below.

We first introduce the Learning with Errors (LWE) problem \cite{regev2005lattices,10.1007/11818175_8} as follows.
   Formally, the assumption is defined by the LWE secret dimension $n \in \mathbb{N}$, the number of available samples $m \in \mathbb{N}$, a modulus $q \ge 2$, and an error distribution $\chi$ over the integers $\mathbb{Z}$. 

\begin{definition}[$(n, m, q, \chi)$-LWE] Distinguish between two distributions:
\begin{itemize}
    \item $(A, As + e)$, where $A \xleftarrow{\$} \mathbb{Z}_q^{m \times n}$, $s \xleftarrow{\$} \mathbb{Z}_q^n$, $e \leftarrow \chi^m$
    \item $(A, r)$, where $A \xleftarrow{\$} \mathbb{Z}_q^{m \times n}$ and $r \xleftarrow{\$} \mathbb{Z}_q^m$
\end{itemize}
\end{definition}

Then we introduce the symmetric-key variant of Regev's LWE-based cryptosystem~\cite{regev2005lattices} as follows, focusing on the descriptions of the $\mathsf{Setup}$, $\mathsf{KeyGen}$, $\mathsf{Enc}$, and $\mathsf{Dec}$ algorithms, as well as the encoding ($\mathsf{Ecd}$) and decoding ($\mathsf{Dcd}$) subroutines.

\begin{itemize}
    \item $pp \leftarrow \mathsf{Setup}(1^\lambda, p)$: Takes the security parameter $\lambda$ and a plaintext modulus $p$ as inputs, and outputs the public parameters $pp$, which consist of the LWE parameters $(n,m, q, \chi)$. All subsequent algorithms implicitly take $pp$ as an implicit parameter.
    
    \item $\bm{sk} \leftarrow \mathsf{KeyGen}(pp)$: Takes the public parameters $pp$ as input and outputs a secret key $\bm{sk} = \bm{s} \xleftarrow{\$} \mathbb{Z}_q^n$.
    
    \item $\bm{\mu'} \leftarrow  \mathsf{Ecd}_{p,q}(\bm{\mu})$: The encoding subroutine parameterized by $p$ and $q$. It takes a plaintext vector $\bm{\mu} \in \mathbb{Z}_p^m$ as input and outputs a vector $\lfloor q/p \rfloor \cdot [\bm{\mu}]_p \in \mathbb{Z}_q^m$.
    
    \item $\mathsf{ct}(\bm{\mu}) \leftarrow \mathsf{Enc}_{\bm{sk}}(\bm{\mu})$: Takes the secret key $\bm{sk} = \bm{s} \in \mathbb{Z}_q^n$ and a message vector $\bm{\mu} \in \mathbb{Z}_p^m$ as inputs. The algorithm samples a matrix $\bm{A} \xleftarrow{\$} \mathbb{Z}_q^{m \times n}$ and an error $\bm{e} \leftarrow \chi^m$. It then outputs the ciphertext $\mathsf{ct}(\bm{\mu}) = (\bm{A}, \bm{A}\bm{s} + \bm{e} + \mathsf{Ecd}_{p,q}(\bm{\mu})) = (\bm{A}, \bm{b})$.
    
    \item $\bm{\mu} \leftarrow \mathsf{Dcd}_{p,q}(\bm{b}')$: The decoding subroutine parameterized by $p$ and $q$. It takes a vector $\bm{b}' \in \mathbb{Z}_q^m$ as input and outputs a plaintext vector $\lfloor \frac{p}{q} \cdot [\bm{b}']_q \rceil \in \mathbb{Z}_p^m$.
    
    \item $\bm{\mu} \leftarrow \mathsf{Dec}_{\bm{sk}}(\mathsf{ct}(\bm{\mu}))$: Takes the secret key $\bm{sk} = \bm{s} \in \mathbb{Z}_q^n$ and a ciphertext $\mathsf{ct}(\bm{\mu}) = (\bm{A}, \bm{b})$ as inputs. It computes $\bm{b}' \leftarrow \bm{b} - \bm{A}\bm{s}$ and outputs the decrypted message vector $\bm{\mu} = \mathsf{Dcd}_{p,q}(\bm{b}')$.
\end{itemize}

It is worth noting that the semantic security of this scheme directly follows from the hardness of the LWE problem, and its correctness is guaranteed as long as $\| \bm{e} \|_\infty < \lfloor q/p \rfloor / 2$.

\begin{remark}[Linear Homomorphism]
     Regev's encryption scheme inherently possesses linear homomorphic properties. Given a ciphertext $\mathsf{ct}(\mu) = (A, b)$ that encrypts a message vector $\mu \in \mathbb{Z}_p^m$, and a linear transformation matrix $D \in \mathbb{Z}^{\ell \times m}$, one can directly evaluate this transformation over the ciphertext~\cite{RLP26}. Specifically, applying $D$ yields $\mathsf{Eval}(D, \mathsf{ct}(\mu)) = (DA, Db)$, which serves as a valid encryption of $D\mu$. It is crucial to note that the successful decryption of this evaluated ciphertext $\mathsf{ct}(D\mu)$ is guaranteed provided that the amplified noise satisfies $\|De\|_\infty < \lfloor q/p \rfloor / 2$, where $e$ represents the initial error term of $\mathsf{ct}(\mu)$.
\end{remark}
\noindent\textbf{Correctness of Regev's LHE.} let $\mathcal{D}_s$ denotes the discrete Gaussian distribution over $\mathbb{Z}$ with variance $s^2/2\pi$. We analysis the the successful decryption probability of above evaluated ciphertext $\mathsf{ct}(D\mu)$. 

\begin{lemma}[Lemma 2.5 in~\cite{CastroL24}]
For any real $s > 0$ and $T > 0$, and any $\mathbf{x} \in \mathbb{R}^n$, we have
\[
\Pr_{\mathbf{e} \leftarrow D_s^m}
\left[
  |\langle \mathbf{x}, \mathbf{e} \rangle|
  \geq T \cdot s \|\mathbf{x}\|_2
\right]
< 2 \exp(-\pi \cdot T^2).
\]
\end{lemma}

For some $\mathbf{D} \in \mathbb{Z}^{\ell \times m}$, consider the decryption of the ciphertext
\(
(\mathbf{A}, \mathbf{A}\mathbf{s} + \Delta \mu + \mathbf{D}\mathbf{e})
\in \mathbb{Z}_q^{m \times (n+1)},
\)
where $\mathbf{e}$ is a fresh sample from $D_s^m$. As noted above, we need
$\Delta/2 > \|\mathbf{D}\mathbf{e}\|_\infty$ to decrypt correctly. To bound
$\|\mathbf{D}\mathbf{e}\|_\infty$, we apply Lemma~2.5 to each row of $\mathbf{D}$. %We can bound the $\ell_2$ norm of any row of $\mathbf{D}$ by $\sqrt{m}\cdot \|\mathbf{D}\|_\infty$. 
Then, setting
\(
T = \frac{\Delta}{2s\|\mathbf{D}[i,\cdot]\|_2},
\)
we have that
\(
\Pr\left[
  \|\mathbf{D}\mathbf{e}\|_\infty \geq \Delta/2
\right] < \delta,
\)
where
\[
\delta =
2 \exp\left(
  -\pi
  \left(
    \frac{\Delta}{2s\|\mathbf{D}[i,\cdot]\|_2}
  \right)^2
\right).
\]

Equivalently, we have decryption failure probability of at most $\delta$ if
\[
\Delta/2
\geq
s\|\mathbf{D}[i,\cdot]\|_2
\cdot
\sqrt{\frac{\ln(2/\delta)}{\pi}}
=
\sigma \cdot \|\mathbf{D}[i,\cdot]\|_2 \sqrt{2\ln(2/\delta)}.
\]
Here, $\sigma$ denotes the standard deviation of $D_s$, i.e.,
$\sigma^2 = s^2/2\pi$. In summary, this gives us a lower bound on $q$ as
\[
q
\geq
2\cdot p \cdot \sigma \cdot \|\mathbf{D}[i,\cdot]\|_2  \sqrt{2\ln(2/\delta)}.
\]
We can bound the $\ell_2$ norm of any row of $\mathbf{D}$ by $\sqrt{m}\cdot \|\mathbf{D}\|_\infty$, which implies that 
\[
q
\geq
2\cdot p \cdot \sigma \cdot \|\mathbf{D}\|_\infty  \sqrt{2m\ln(2/\delta)}.
\]
%This saves us roughly a $\sqrt{m}$ factor over the naive worst-case analysis.

\subsection{Private Information Retrieval}
{Private Information Retrieval (PIR)} enables a client to privately query a remote database to retrieve a desired item~\cite{646125}. A PIR scheme with preprocessing consists of two phases: a one-time preprocessing phase, during which the server preprocesses the database and outputs reusable hints, thereby significantly reducing the overhead during the query phase; and a query phase, during which the client queries the server for a specific entry of the database \textit{without revealing the query itself}.\medskip

\noindent \textbf{PIR from LHE.} A natural approach to constructing PIR from LHE \cite{RLP26} involves reshaping the database into a matrix representation. Concretely, a database of $N$ items is organized into an $l \times m$ matrix $\bm{D}$ such that $l \times m \ge N$. To retrieve the $i$-th item, the client computes its corresponding row index $i_r \in [l]$ and column index $i_c \in [m]$. It then generates an LHE encryption of the $i_c$-th unit vector of size $m$ and transmits it to the server. The server treats $\bm{D}$ as a linear transformation, evaluates it over the ciphertext, and returns the response. The client concludes the protocol by decrypting the server's response and extracting the target item using the row index $i_r$.\medskip

\noindent \textbf{SimplePIR.} While it is possible to construct a PIR scheme using LHE, directly applying the well-known Regev's cryptosystem presents significant practical challenges. Specifically, Regev encryption entails a public random matrix $A \in \mathbb{Z}_q^{m \times n}$. Homomorphic matrix multiplication requires the server to compute the massive matrix product $DA$ to process the query. Given the strict parameterizations required to preserve LWE hardness, computing $DA$ dynamically during the online phase results in prohibitive computational and communication overheads.

SimplePIR~\cite{10.5555/3620237.3620455} resolves this bottleneck by observing that the computationally dominant matrix product $DA$ is entirely independent of the online query. By exploiting this structural independence, SimplePIR decouples the cryptographic workload, shifting the heavy computation to an offline preprocessing phase:

\begin{itemize}
    \item Offline Preprocessing: During a one-time setup, the server and client agree on a public LWE matrix $A \in \mathbb{Z}_q^{m \times n}$. The server computes $H = DA \pmod q$ and transmits this massive matrix to the client as a cryptographic "hint," which the client persistently stores.
    \item Online Query: To retrieve a record via a selection vector $\mu$, the client samples a fresh secret $s \in \mathbb{Z}_q^n$ and an error vector $e$ to guarantee semantic security, sending the query ciphertext $u = As + e + \text{Ecd}(\mu) \pmod q$. The server processes this query using only a lightweight matrix-vector multiplication, returning $v = Du \pmod q$.
    \item Client Recovery: The client decrypts the response using its stored hint $H$ and ephemeral secret $s$. By computing $v - Hs$, the term reduces to $De + D\cdot\text{Ecd}(\mu)$. Provided the amplified noise $\|De\|_\infty < \lfloor q/p \rfloor / 2$, the client successfully extracts the target item $D\mu$.
\end{itemize}
By shifting the expensive $DA$ computation entirely offline, SimplePIR reduces the server's online workload to a single matrix-vector multiplication. While numerous subsequent works have explored various vertical optimizations—such as eliminating client-side storage constraints (e.g., Hintless PIR) or optimizing client-side compute—SimplePIR's architectural paradigm remains the definitive state-of-the-art benchmark for \emph{online performance} in semi-honest PIR.

\subsection{Verifiable Private Information Retrieval}

In contrast to standard PIR, in a verifiable PIR scheme, the server is additionally required to set up a digest of the database during the preprocessing phase. During the query phase, besides ensuring the privacy of the client's query, we also require the integrity of the server's response—namely, that the returned data is consistent with the database bound to the digest.
Following \cite{CastroL24,RLP26}, a vPIR scheme consists of a public parameter setup $pp\gets \textsf{Setup}(1^\lambda,1^N,p)$, where $N$ is the database size and $p$ is the database entry modulus, and the following three phases: 
\begin{itemize}
    \item \textbf{One-Time Setup phase:} This phase is independent of the client. The server generates a digest of its database as \(d \gets \textsf{Digest}(pp, D)\). The validity of the digest can be publicly verified via \(b_{dig} \gets \textsf{DigVer}(pp, d)\), where $b_{dig}$ is a bit indicating acceptance or rejection. Notably, this digest also serves as a hint.
    
    \item \textbf{Proof Preprocessing phase:} 
    This phase is a protocol \(\langle S_\text{pre}(D), C_\text{pre} \rangle(pp, d)\) between the server and the client. Both parties take the parameter \(pp\) and the digest \(d\) as input, and the server additionally takes the database \(D\) as input. At the end of the protocol, the client outputs a bit \(b_{pre} \in \{0,1\}\) indicating acceptance or rejection. Upon acceptance, the client stores a reusable state \(\textsf{st}_{pre}\) locally for use in the query phase, and proceeds to the query phase only in that case.
    \item \textbf{Query phase:} This phase is a protocol \(\langle S_{\text{query}}(D), C_{\text{query}}(\textsf{st}_\text{pre}, i) \rangle(pp, d)\) between the server and the client. The client holds a reusable state \(\textsf{st}_\text{pre}\) and a desired index \(i\). The server holds the database \(D\). At the end of the protocol, the client outputs a pair \((b_{\text{query}}, a)\), where \(b_{\text{query}} \in \{0,1\}\) indicates acceptance or rejection, and if \(b_{\text{query}} = 1\), then answer \(a = D[i]\) (the retrieved entry); otherwise \(a = \bot\). %Correctness requires that for an honest server, the client outputs \(D[i]\) with overwhelming probability; security requires that the server learns no information about \(i\), and soundness guarantees that if the server deviates, the client rejects except with negligible probability.
\end{itemize}

A secure vPIR scheme must satisfy the following requirements:

\begin{itemize}
    \item \textbf{Completeness:} If the server and client follow the protocol faithfully, then with overwhelming probability the client accepts in all phases and retrieves the correct database entry $D[i]$.
    \item \textbf{Digest Binding:} The digest \(d\) satisfies two requirements:

\begin{enumerate}
    \item \textbf{Computationally binding.} The digest \(d\) constitutes a \textit{computationally binding commitment} to the underlying database \(D\). %That is, no efficient adversary can find two distinct databases \(D \neq D'\) that both map to the same digest under the digest algorithm.

    \item \textbf{Extractability.} There exists a PPT oracle extractor \(\mathcal{E}\) such that for any PPT malicious server \(S^*\) that outputs a digest \(d\), the following holds: the probability that \(\textsf{DigVer}(pp,d)=1\) while the extracted pair \((D,\tau) \gets \mathcal{E}^{S^*}(pp,d)\) is \textbf{not} a valid decommitment of \(d\) to \(D\) is negligible. Here the probability is taken over the randomness of the setup, the adversary \(S^*\), and the extractor \(\mathcal{E}\)
\end{enumerate}
    
    %Digest binding property consists of two part: 1) The digest $d$ constitutes a \textit{computationally binding commitment} to the underlying database $\mathbf{D}$; and 2) There exists a PPT oracle extractor $\mathcal{E}$ satisfying that, for any PPT malicious $S^*$ 
    %$$\Pr[\textsf{DigVer}(pp,d)=1\land \tau \mbox{ is not a valid decommitment for } d \mbox{ to }D|pp\gets \textsf{Setup}(1^{\lambda},1^N,p),d\gets S^*(pp),(D,\tau)\gets \mathcal{E}^{S^*}(pp,d)]$$

\item \textbf{Query Hiding:}  This property requires that a malicious PPT server cannot learn any information about the client's query from the interaction, even if the server is allowed to observe whether the client accepts the query phase. Formally, we require the existence of a PPT simulator \(\mathsf{Sim}\) such that for any polynomial-size malicious server \(S^*\), for any sequence of polynomially many queries \(\tau=(i_1,\cdots, i_k)\), and for random public parameter $pp\gets \textsf{Setup}(1^\lambda,1^N,p)$, the following holds:
$$\textsf{QH}^{S^*}(pp,\tau)\stackrel{c}{\approx} \textsf{Sim}(S^*,pp) $$
where $\textsf{QH}^{S^*}(pp,\tau)$ is a random variable describing \textit{the view of $S^*$} in the following experiment: 
\begin{enumerate}
    \item Given public parameter $pp$, \(S^*\) outputs a digest \(d\) such that \(\textsf{DigVer}(pp,d)=1\).
    \item \(S^*\) then interacts with an honest client \(C\) in the proof preprocessing phase \(\langle S^*, C_{\textsf{pre}} \rangle(pp,d)\). Let \(b_{\textsf{pre}}\) be the client's output. If \(b_{\textsf{pre}} = 0\), the experiment aborts.
    \item For each query \(i_j\) in \(\tau\) (in order), \(S^*\) interacts with \(C\) in the query phase \(\langle S^*, C_{\textsf{query}}(\textsf{st}_{\textsf{pre}}, i_j) \rangle(pp,d)\). The adversary \(S^*\) is allowed to see \textbf{the acceptance bit \(b_{\textsf{query}}^{(j)}\)} after each phase (i.e., whether the client accepted the query response).
\end{enumerate}
    
    \item \textbf{Soundness:} 
    %It is computationally infeasible for a malicious server to convince the client to accept an inconsistent answer \(a \neq D[i]\), where \(D\) is the database committed to in the digest \(d\). More formalized, for any malicious server $S^*$ and any polynomial queries $i_1,i_2,\cdot, i_k$ (even chosen adaptively by $S^*$), 
A vPIR scheme is sound if it is computationally infeasible for a malicious server to convince the client to accept an incorrect answer  \(a \neq D[i]\) for a query $i$. More formally, for any PPT malicious server $S^*$ and any polynomial number of queries $i_1,\dots,i_k$ (which may be chosen adaptively by $S^*$), the probability that $S^*$ wins the following game $\mathsf{G}^{\mathsf{soundness}}_{S^*}(\lambda)$ is negligible in the security parameter $\lambda$:\medskip

\noindent
\fbox{\parbox{\linewidth}{%
\vspace{2pt}{$\mathsf{G}^{\mathsf{soundness}}_{S^*}(\lambda)$:}
\begin{enumerate}[leftmargin=2em, labelindent=0pt, itemindent=0pt]
    \item \(pp \gets \textsf{Setup}(1^\lambda, 1^N, p)\).
    \item \(S^*\) outputs a digest \(d\) such that \(\textsf{DigVer}(pp,d)=1\).
    \item \(S^*\) interacts with an honest client \(C\) in the proof preprocessing phase \(\langle S^*, C_{\textsf{pre}} \rangle(pp,d)\). The client outputs \(b_{\textsf{pre}}\). If \(b_{\textsf{pre}}=0\), the game aborts.
    \item \(S^*\) then interacts with \(C\) in a sequence of query phases. For each query \(i_j\), $S^*$ and $C$ run \(\langle S^*, C_{\textsf{query}}(\textsf{st}_{\textsf{pre}}, i_j) \rangle(pp,d)\) and $C$ obtains \((b_{\textsf{query}}^{(j)}, a_j)\). \(S^*\) is allowed to learn \(b_{\textsf{query}}^{(j)}\) immediately after the \(j\)-th query phase.
\end{enumerate}
\textbf{Win condition.} The adversary wins if there exists \(j\) such that \(b_{\textsf{query}}^{(j)} = 1\) yet \(a_j \neq D[i_j]\), where \(D\) is the database extracted from \(d\) via the extractor guaranteed by the digest binding property.
\vspace{2pt}
}}\vspace{1em}

%The probability is taken over the randomness of $\textsf{Setup}$, the adversary $S^*$, the honest client, and the extractor.
    
    %produce an answer that passes $\mathsf{Verify}$, unless it is the exact, unique valid answer corresponding to the query $q$ and the committed database $\mathbf{D}$.
\end{itemize}

\section{Database Commitment with Linear Relation Proofs and Range Proofs}
\subsection{Error-Correcting Code--Based commitment}

Instead of using \( H = DA \) as a commitment, which relies on SIS assumptions, we adopt a simpler method to generate a commitment to \( D \) and use it as the digest of \( D \). That is, we view \( D \) as a matrix, apply error-correcting code encoding row by row, and finally hash the resulting encoded matrix. At this point, the hash actually constitutes a polynomial commitment to a polynomial determined by \( D \). Therefore, in the following subsections, we verify the ``well-formedness'' of the commitment and further prove linear relations among its rows based on the tensor-IOP scheme from Ligero/Brakedown\cite{AHIV17,GLSTW23}. In this subsection, we first provide the description of the commitment and decommitment algorithms. \medskip

\noindent\textbf{Commitment.} Let $q$ be a prime{\iffalse(in the following construction, we would set $q$ the ciphertext modulus of the LHE encryption scheme)\fi}, let $D \in \mathbb{Z}_p^{\ell \times m}$ be a database and let $D[i, \cdot]$ denote its $i$-th row. Denote by $RS[\mathbb{F}_q, m, \rho]$ the Reed–Solomon code over $\mathbb{F}_q$ with rate $\rho$, which maps a string $s \in \mathbb{F}_q^\ell$ to a codeword $RS(s) \in \mathbb{F}_q^{m \rho^{-1} }$. The committer encodes each row of $D$ using RS and then hashes each column of the encoded matrix $\hat{D}$ to obtain the commitment.\smallskip
%Let $D\in \mathbb{Z}^{\ell\times m}$ be a database, $D[\cdot,j]$ the $j$-th column of $D$, and $q$ the ciphertext module of LHE encryption. Denote by $RS[\mathbb{F}_q,\ell,\rho]$ the Reed-Solomon code over $\mathbb{F}_q$ which on input a string $r\in \mathbb{F}^{m}_{q}$, outputs the code $RS(r)\in \mathbb{F}^{\rho^{-1}m}_q$, where $\rho$ is the rate of RS code.

\noindent\textbf{Decommitment.} The decommitment phase, parameterized by $\beta\le \frac{1-\rho}{2}$, is straightforward: the committer sends the database $D$, the encoded matrix $\hat{D}$, and used randomness as the decommitment. The receiver accepts if 1) the commitment is indeed the hash values of columns of $\hat{D}$, and 2) for each $i\in [\ell]$, $\hat{D}[i,\cdot]$ is $\beta$-close to the encoding of $D[\cdot, j]$, i.e., the relative distance between $\hat{D}[i,\cdot]$ and the encoding of $D[i, \cdot]$ is at most $\beta$.\smallskip

A formal description of the commitment and decommitment phases is presented in Figure \ref{fig.com}.\smallskip

%\begin{figure}
%	\begin{algorithmic}
%    \algrule{1.2pt}
		%\normalsize
%       \STATE\textbf{Commit:} On input a database $D \in \mathbb{Z}^{\ell \times m}$, the committer encodes each row $D[i,\cdot]$ for $i \in [\ell]$ using the Reed–Solomon\cite{doi:10.1137/0108018} code $RS[\mathbb{F}_q, m, \rho]$, obtaining an encoded matrix $\hat{D}$. It then hashes each column of $\hat{D}$ to obtain a vector of hash values $h_D=(h_0,\cdots,h_{m\rho^{-1}-1})$. Set commitment as $\textsf{com}=h_D$. All randomness used during the commitment phase, along with $D$ and $\hat{D}$, constitutes the decommitment $\tau$. \smallskip
        
%      \STATE\textbf{Decommit:} On input the decommitment $\tau$, the receiver accepts the decommitment if and only if: 1) $h_{D}$ is the hash values of columns of $\hat{D}$, and 2)  for each $i\in[\ell]$, $$\Delta(\hat{D}(i,\cdot), \textsf{RS.Encode}(D(i,\cdot)))\le\beta\le \frac{1-\rho}{2},$$ where $\rho$ is the rate of RS code, $\beta$ is a scheme parameter.
%     \algrule{0.4pt}
%	\end{algorithmic}
%\caption{Commitment and decommitment algorithm}
%	\label{fig.com}
%\end{figure}

\begin{figure}
\algrule{1.2pt}
\algrule{0.4pt}
	\begin{description}[leftmargin=0pt]
    \item[\textbf{Commit:}] On input a database $D \in \mathbb{Z}^{\ell \times m}$, the committer encodes each row $D[i,\cdot]$ for $i \in [\ell]$ using the Reed–Solomon\cite{doi:10.1137/0108018} code $RS[\mathbb{F}_q, m, \rho]$, obtaining an encoded matrix $\hat{D}$. It then hashes each column of $\hat{D}$ to obtain a vector of hash values $h_D=(h_0,\cdots,h_{m\rho^{-1}-1})$. Set commitment as $\textsf{com}=h_D$. All randomness used during the commitment phase, along with $D$ and $\hat{D}$, constitutes the decommitment $\tau$.\medskip
    \item[\textbf{Decommit:}] On input the decommitment $\tau$, the receiver accepts the decommitment if and only if: 1) $h_{D}$ is the hash values of columns of $\hat{D}$, and 2)  for each $i\in[\ell]$, $$\Delta(\hat{D}(i,\cdot), \textsf{RS.Encode}(D(i,\cdot)))\le\beta\le \frac{1-\rho}{2},$$ where $\rho$ is the rate of RS code, $\beta$ is a scheme parameter.
	\end{description}
    \algrule{0.4pt}
\caption{Commitment and decommitment algorithm}
	\label{fig.com}
\end{figure}

\begin{lemma} The commitment scheme provided in Figure \ref{fig.com} satisfying  the completeness and binding property.
\end{lemma}
\noindent\textit{Proof.}  Completeness follows directly from our construction. Computational binding relies on the collision resistance of the hash function and the properties of the RS code. To open a commitment to two distinct databases, an adversary must either provide two different encoded matrices $\hat{D}$ and $\hat{D}'$ that hash to the same commitment, thereby breaking collision resistance, or provide the same $\hat{D}$ but two different databases $D$ and $D'$ with $D \neq D'$. The latter is also impossible, since there is at most one $D$ whose encoding has relative distance from $\hat{D}$ less than $\beta\le\frac{1 - \rho}{2}$. Recall that $\frac{1 - \rho}{2}$ is the unique decoding distance of the RS encoding scheme.\qed

\subsection{Linear relation proof}
%本小节中，我们基于Ligero/Brakedown中的张量IOP协议构造对被承诺的D的列线性关系的证明。这个证明包括两个部分：第一部分是一个测试算法，该算法用来证明承诺的Well-formedness(i.e,存在一个合法的decommitment), 第二部分是线性关系证明算法。其中第一部分只需要执行一次，即便我们需要证明多个线性关系。
In this subsection, we construct a proof for the row linear relations of a committed $D$ based on the tensor IOP protocol from Ligero/Brakedown. The proof consists of two parts: the first part is a testing protocol that proves the well-formedness of the commitment (i.e., that a valid decommitment exists), and the second part is the linear relation proving protocol. Notably, the first part needs to be executed only once, even when multiple linear relations are to be proved.\medskip

%The proof of column linear relations consists of two phases: a test phase, which is used to prove the well-formedness of the commitment and is executed only once (even when multiple linear relation proofs are required), and a proof phase.\smallskip

\noindent\textbf{Testing protocol.} Our testing protocol mirrors the testing phase of Brakedown's polynomial commitment scheme, with the exception that we refine the parameter settings to flexibly trade off between the overhead of the testing protocol (which lies in the setup phase of our vPIR construction) and the overhead of the proving protocol (which might lie in the online phase).

On input the commitment $\textsf{com}$, the verifier first sends a random vector $r = (r_0,r_1,\cdots,r_{\ell-1})$, where each $r_i$ is sampled uniformly from $\mathbb{F}_{q^t}$ and $t$ is an integer chosen such that $1/q^t$ is negligible. The prover then responds with the linear combination $\alpha$ of the rows $D[i,\cdot]$ using the scalars $r_i$, i.e., $\alpha=\sum_{i \in [\ell]} r_i \cdot D[i,\cdot]$.\footnote{The multiplication $r_i \cdot D[i,\cdot,]$ requires first viewing each entry of $D$ (which lies in $\mathbb{F}_q$) as an element of $\mathbb{F}_{q^t}$, and then performing multiplication in $\mathbb{F}_{q^t}$. A Reed–Solomon code over \(\mathbb{F}_q\) can be naturally extended to \(\mathbb{F}_{q^t}\) by encoding each component separately; under this extension, any codeword originally over \(\mathbb{F}_q\) remains a valid codeword over \(\mathbb{F}_{q^t}\).}

Note that if there exists a row $\hat{D}[i,\cdot]$ which is far from codewords, then, with overwhelming probability, the linear combination $\sum_{i \in [\ell]} r_i \cdot \hat{D}[i,\cdot]$ will also be far from the encoding of $\sum_{i \in [\ell]} r_i \cdot {D}[i,\cdot]$. Consequently, the verifier selects a random set $I\subset [m\rho^{-1}]$ of size $Q_t=\Omega(-\lambda\cdot\log_{1-\beta}2)$, and the prover retrieves $\{\hat{D}[\cdot,j]\}_{j\in I}$ by opening the hash values $\{h_{j}\}_{j\in I}$.

The verifier checks that for each $j \in I$: (1) the received $\hat{D}[\cdot,j]$ indeed hashes to $h_i$, and (2) the inner product of the received $\hat{D}[\cdot,j]$ with the vector $r$ is consistent with the $j$-th element of the encoding of $\alpha$. This ensures that the rows of $\hat{D}$ are close to codewords.

The formalized construction is shown in Figure \ref{fig.test}.
\begin{figure}
	%\begin{algorithmic}
    \algrule{1.2pt}
    \algrule{0.4pt}\smallskip   
		\normalsize
        \textbf{Testing protocol:}\smallskip       
        \begin{enumerate}
            \item The verifier chooses and sends a random vector $r=(r_0,r_1,\cdots,r_{\ell-1})\in \mathbb{F}^{\ell}_{q^t}$ to the prover.
            \item The prover computes and sends the linear combination of the columns $D[i,\cdot]$ with scalars $\{r_i\}$, i.e., $\alpha=\sum_{i \in [\ell]} r_i \cdot D[i,\cdot]$.  
            \item The verifier computes the encoding $RS(\alpha)$ over $\mathbb{F}_{q^t}$. It chooses a random subset $I$ of $[m\rho^{-1}]$ with size $Q_t=\Omega(-\lambda\cdot\log_{1-\beta}{2})$ and sends it to the prover.
            \item For each $j \in I$, the prover sends the $j$-th column of $\hat{D}$, i.e., $\hat{D}[\cdot,j]$, to the verifier, along with the opening (used randomness) of hash values $h_j$.
            \item The verifier checks that for each $j \in I$: 
            \begin{itemize}
                \item the received $\hat{D}[\cdot,j]$ indeed hashes to $h_j$; 
                \item the inner product of column $\hat{D}[\cdot,j]$ and vector $r$ is consistent with the $j$-th element of $RS(\alpha)$.
            \end{itemize}
            The verifier accepts if all checks pass.
        \end{enumerate} 
         \algrule{0.4pt}
	%\end{algorithmic}
\caption{Testing protocol}
	\label{fig.test}
\end{figure}
\begin{lemma}\label{the.pcs1}
    The protocol described in Figure~\ref{fig.test} is an argument of knowledge for the statement that there exists a valid decommitment to the commitment $\textsf{com}$.
\end{lemma}
\noindent\textit{proof.} The \textbf{Completeness} is obvious.

In the proof of \textbf{Knowledge Soundness}, we model the hash function as a random oracle, which allows us to view the protocol as an IOP system. %We show that a malicious prover cannot pass the verifier's checks with non-negligible probability. 
As discussed in~\cite{GLSTW23} and in the existing
IOP-to-succinct-argument transformation of~\cite{DBLP:conf/tcc/Ben-SassonCS16}, given any prover that convinces the argument verifier to accept with non-negligible probability, there exists an efficient straight-line extractor that outputs an IOP proof string $\pi$ which ``opens'' the hash commitment sent by the argument-system prover in the commitment phase. Recall that the encoding matrix $\hat{D}$ is hashed in our construction, therefore the extractor can extract $\hat{D}$  together with the randomness used for hashing, and then decode each row within the unique decoding radius of the code to retrieve database $D$. 

 Next, we show that if the malicious prover can pass the testing protocol with noticeable probability, then the decommitment and database extracted above is available.  To this end, we first recall the correlated agreement theorem for Reed–Solomon codes.
\begin{theorem}[Correlated agreement over affine spaces~\cite{DBLP:conf/focs/Ben-SassonCIKS20}]\label{cor. agg} Let $V:=\textsf{RS}[\mathbb{F},m,\rho]$ be a Reed–Solomon code of length $m\rho^{-1}$ and rate $\rho$. For any $u_0,u_1,\cdots,u_{\ell}\in \mathbb{F}^{m\rho^{-1}}$, let $U=u_0+\textsf{span}\{u_1,\cdots,u_{\ell}\}\subset\mathbb{F}^{m\rho^{-1}}$ be an affine subspace. If $\beta\in (0,\frac{1-\rho}{2}]$ and 
$$\Pr_{u\in U}[\Delta(u,V)\le \beta]>\epsilon,$$
where $\epsilon=\frac{m\rho^{-1}}{|\mathbb{F}|}$. 
Then there exist a set $J\subset [m\rho^{-1}]$ and codewords $v_0,v_1,\cdots,v_\ell \in V$ satisfying:
\begin{itemize}
    \item \textbf{Density:} $|J|/m\rho^{-1}\ge 1-\beta$, and 
    \item \textbf{Agreement:} for all $i\in \{0,1,\cdots,\ell\},$ the strings $u_i$ and $v_i$ agree on all coordinates in $J$.
\end{itemize}
\end{theorem}

Now we claim that, if the prover passes the verifier's checks with probability greater than
\(\epsilon + (1-\beta)^{|Q_t|}\), then there must exist a sequence of
\(\ell\) Reed--Solomon codewords \(c_1,\ldots,c_\ell\) such that\medskip
%\[
%E:=\left|\left\{j\in[m\rho^{-1}]:\exists i\in [\ell], \text{ such that }c_{i,j}\neq \hat{D}[i,j]\right\}\right|\le \beta m\rho^{-1},
%\]

\noindent\resizebox{\linewidth}{!}{$
    \displaystyle
    E:=\left|\left\{j\in[m\rho^{-1}]:\exists i\in[\ell],\text{ such that }c_{i,j}\neq \hat{D}[i,j]\right\}\right| \le \beta m\rho^{-1}
$}\medskip

\noindent which means that the extracted $\hat{D}$ is decodable, and $\hat{D}$ together with hashing randomness constitutes a valid decommitment.

Assume, for contradiction, that the condition does \textbf{not} hold. This means that no set \(J\) satisfying the density and agreement requirements of Theorem~\ref{cor. agg} exists.
%Let $u = \sum_{i=1}^{\ell}r_i\cdot \hat{D}[i,\cdot]$. By
%Theorem~\ref{cor. agg}, the probability, over the verifier's choice of $r$,
%that there exists a Reed--Solomon codeword $c$ satisfying \(\Delta(u,c)\le\beta\) is at most \(\epsilon\). If no such $c$ exists, then
%\(\Delta(u, \textsf{RS.Encode}(\alpha))>\beta\). In this event, all of the
%verifier's consistency tests pass with probability at most
%\((1-\beta)^{|Q_t|}\). Therefore, the prover passes the verifier's checks with probability less than \(\epsilon + (1-\beta)^{|Q_t|}\), toward a contradiction.
Consider the linear combination \(u = \sum_{i=1}^{\ell} r_i \cdot \hat{D}[i,\cdot]\) for random coefficients \(r_i\) chosen by the verifier. 
By Theorem~\ref{cor. agg}, the probability (over the verifier’s choice of \(r\)) that there exists a Reed–Solomon codeword \(c\) with \(\Delta(u,c) \le \beta\) is at most \(\varepsilon\). 
If no such codeword exists, then \(\Delta(u, \textsf{RS.Encode}(\alpha)) > \beta\) for any message \(\alpha\). 
In this case, the verifier’s consistency tests pass with probability at most \((1-\beta)^{|Q_t|}\). 
Consequently, the prover’s overall probability of passing the verifier’s checks is bounded by \(\varepsilon + (1-\beta)^{|Q_t|}\), contradicting our initial assumption.

Since \(\beta \in (0, \frac{1-\rho}{2}]\), we have \((1-\beta)^{|\Omega(-\lambda\cdot\log_{1-\beta}2)|} \le 2^{-\lambda}\), which is negligible. This completes the proof of knowledge soundness.
%Since $\delta\in (0,\frac{1-\rho}{2}]$,  the probability \((1-\beta)^{|\Omega(-\lambda\cdot\log_{1-\beta}2)|}\le 2^{-\lambda}\) is negligible, the proof is conclude.
\qed\medskip

\noindent\textbf{Proving protocol.} For a commitment $\textsf{com}$ that has been proven to be well-formed using \textsf{Test} protocol, the following proving protocol allows us to prove linear relations among the rows of the committed database $D$, e.g., $\alpha = \sum_{i \in [\ell]} a_i \cdot D[i,\cdot]$.

The proof protocol can be viewed as a simplified version of the testing protocol. The difference is that the random vector $r$ that the verifier originally sends in the first round is now fixed to $a = (a_1, \dots, a_\ell)$. The value $\alpha$ that was originally sent in the second round becomes exactly the linear relation to be proved, i.e., $\alpha = \sum_{i \in [\ell]} a_i \cdot D[i,\cdot]$. The subsequent steps, i.e., the selection of the random challenge subset $I$ and the corresponding responses, remain the same as in the testing protocol, except that the size of $I$ is set to $\Omega(-\lambda \cdot \log_{\rho+\beta} 2)$.\smallskip

The formalized construction is shown in Figure \ref{fig.linearproof}.
\begin{figure}
	%\begin{algorithmic}[H]
    \algrule{1.2pt}
    \algrule{0.4pt}\smallskip  
		\normalsize
        \textbf{Proving protocol:} For a commitment $\textsf{com}$ that has been proven well-formed using \textsf{Testing} protocol and a vector $\alpha$, the prover wants to prove that the committed $D$ satisfying $\alpha = \sum_{i \in [\ell]} a_i \cdot D[i,\cdot]$. Both parties run as follows:\smallskip     
        \begin{enumerate}
            \item The verifier computes the encoding $RS(\alpha)$. It chooses a random subset $I$ of $[m\rho^{-1}]$ with size $\Omega(-\lambda \cdot \log_{\rho+\beta} 2)$ and sends it to the prover.
            \item For each $j \in I$, the prover sends the $j$-th column of $\hat{D}$, i.e., $\hat{D}[\cdot,j]$, to the verifier, along with the opening (used randomness) of hash values $h_j$.
            \item The verifier checks that for each $j \in I$: 
            \begin{itemize}
                \item the received $\hat{D}[\cdot,j]$ indeed hashes to $h_j$; 
                \item the inner product of column $\hat{D}[i,\cdot]$ and vector $a$ is consistent with the $j$-th element of $RS(\alpha)$.
            \end{itemize}
            The verifier accepts if all checks pass.
        \end{enumerate} 
    \algrule{0.4pt} 
	%\end{algorithmic}
\caption{Proving algorithm}
	\label{fig.linearproof}
\end{figure}
\begin{lemma}\label{the.pcs}
    The protocol described in Figure~\ref{fig.linearproof} is an argument of knowledge for the statement that $\alpha = \sum_{i \in [\ell]} a_i \cdot D[i,\cdot]$ where $D$ is the database committed in $\textsf{com}$.
\end{lemma}
\noindent\textit{proof.}
Recall from the proof of the testing protocol, if the prover can prove the well-formedness of commitment, the extractor can extract $\hat{D}$ and decode it to obtain the committed $D$. We now show that if the prover can pass the checks in the proving protocol with probability greater than \(\left(\rho+\beta\right)^{|I|}\), then it must hold that $\alpha  = \sum_{i \in [\ell]} a_i \cdot D[i,\cdot]$. 

%that \(c_i\) denotes the Reed--Solomon codeword closest to the \(i\)-th row of \(\hat{D}\), and let \(w:=\sum_{i=1}^{\ell} a_i\cdot c_i\). We next show that if the prover passes the testing protocol with probability greater than \(\epsilon+(1-\beta)^{|Q_t|}\), and also passes the checks in the proving algorithm with probability greater than \(\left(1-\rho-\beta\right)^{|I|}\), then it must hold that \(w=\textsf{RS.Encode}\left(\sum_{i=1}^{\ell} a_i\cdot D[i,\cdot]\right)\).
 Otherwise, toward a contradiction, assume that \(\alpha\neq \sum_{i=1}^{\ell} a_i\cdot D[i,\cdot]\). Then \(w=\textsf{RS.Encode}\left(\sum_{i=1}^{\ell} a_i\cdot D[i,\cdot]\right)\) and \(\textsf{RS.Encode}(\alpha)\) are two distinct Reed--Solomon codewords,
%Let \(\alpha=\sum_{i=1}^{\ell} a_i\cdot D[i,\cdot]\). Suppose, toward a contradiction, that \(w\neq \textsf{RS.Encode}(\alpha)\). Then \(w\) and \(\textsf{RS.Encode}(\alpha)\) are two distinct Reed--Solomon codewords, 
and therefore they agree on at most \(m\) positions. 
 Let \(A\) denote the set of
positions on which they agree. In the proving algorithm, the verifier rejects whenever there exists some \(j\in I\) such that \(j\notin  A\cup E\), where \(E\) is the set defined in the proof of Lemma~\ref{the.pcs1}. Since
\(|A\cup E|\le |A|+|E|\le m+\beta m\rho^{-1}
= (\rho+\beta)m\rho^{-1}\), a uniformly random column
\(j\in[m\rho^{-1}]\) lies in \(A\cup E\) with probability at most
\(\rho+\beta\). It follows that the prover passes the checks in the proving
algorithm with probability at most \(\left(\rho+\beta\right)^{|I|}\). 

Again, since $\beta\in (0,\frac{1-\rho}{2}]$, we have \(\left(\rho+\beta\right)^{\Omega(-\lambda\cdot\log_{\rho+\beta}2)}\le 2^{-\lambda}\) is negligible. This completes the proof. \qed

%Therefore, if the prover cheats, it can pass both the testing protocol and the checks in the proving algorithm only with negligible probability. As in the discussion of knowledge soundness in Lemma~\ref{the.pcs1}, we model the hash function as a random oracle. Hence, in the random oracle model, the proving algorithm satisfies knowledge soundness.
%\qed

\begin{remark} An interesting observation is that, in the above proof, even if the prover does not know the values of $a_i$ and $\alpha$, it does not affect the execution of the proof protocol. This is because the prover only needs to open certain hash values specified by the verifier. In the construction of our final vPIR protocol, we similarly exploit this property to complete the proof of $Z = CD$, where the prover does not know $C$ and $Z$, but only knows the encryption of $C$.
\end{remark}

\noindent\textbf{Batched Proving Algorithm.} The above protocol can be directly extended to prove multiple row linear relation claims $\{\alpha_k = \sum_{i \in [\ell]} a_{k,i} \cdot D[i,\cdot]\}_k$. The rows of $\hat{D}$ opened in the first two steps of the algorithm are reused to verify all of the above claims, which does not affect the soundness analysis: for any false claim, the probability that the proof passes verification remains the same as before. %The construction of the scheme is as follows:

%\noindent\textit{Scheme Description.} This scheme is identical to the one in Figure~3, except that in the third step (the verifier's verification phase), the verifier checks each claim: for each $k$, the inner product of row $\hat{D}[i,\cdot]$ with the vector $a_k = (a_{k,1}, \dots, a_{k,m})$ is consistent with the $i$-th element of the encoding of $\alpha_k$.

\subsection{Range Proof}
In this subsection, we present two new approaches to range proofs. The first method is an efficient almost precise range proof. Unlike previous schemes, it proves the Euclidean norm of the database, and its result is significantly stronger: the loss in parameter selection for the plaintext modulus is less than 1 bits, whereas previous schemes incur a loss of up to \(0.5\log \ell\) bits (7–9 bits). The second method is an precise range proof, which is relatively less efficient, but this process only occurs during the setup phase.

More specifically, the first approach is based on the Johnson–Lindenstrauss lemma; it uses specific random challenges and checks the Euclidean norm of the response to ensure that the Euclidean norm of the committed value does not deviate too much. The second approach is based on SNARK techniques, where the commitment to $D$ is viewed as a polynomial commitment to some polynomial $f_D$ associated with $D$. Using lookup arguments in SNARKs, we reduce the range check to polynomial commitment evaluations (i.e., linear relation checks), thereby proving the range of $D$ precisely.\medskip

\noindent \textit{First approach.} This method follows from the famous Johnson-Lindenstrauss Lemma: any set of points in a high-dimensional space can be projected into a much lower-dimensional space while approximately preserving all pairwise Euclidean distances within a \textit{constant} factor.

%if the infinity norm of $D$ is too large, then for a random  column linear combination with coefficients chosen from $\{0,1\}$, the resulting will, with sufficiently high probability, exceed its expected range. This observation has been used not only in previous SIS commitment schemes but also widely applied in range checks involving unknown-order groups and lattices. 

This approach allows us to directly prove the Euclidean norm of each row of database $D$, thereby ensuring that the decryption failure probability of Regev encryption in the online phase is sufficiently small. Specifically, to prove that $\|D[i,\cdot]\|_2 \le b\sqrt{m}$ for each row $D[i,\cdot]$ (suppose that $\|D\|_\infty \le b$) for $D \in \mathbb{Z}^{\ell \times m}$, the verifier picks a random binary matrix $\Pi \gets \mathcal{D}^{m \times 256}$ and sends it to the prover, where $\mathcal{D}$ is the distribution that $\Pr[0\gets \mathcal{D}]=0.5$ and $\Pr[-1\gets \mathcal{D}]=\Pr[1\gets \mathcal{D}]=0.25$ . The prover computes and sends $D^* = D\cdot \Pi \mod{q}$ to the verifier\footnote{For an element $w \in \mathbb{F}_q$, we here take its representative in the interval $[-q/2, q/2]$.}, who checks that $\|D^*[i,\cdot]\|_2 \le b\sqrt{337m}$ for each $i$. The verifier then randomly selects and sends a random matrix $C \in \{0,1\}^{\lambda \times \ell}$, and the prover responds $Z=CD$. Both parties using the row linear relation proof protocol to verify that $Z=CD \mod q$ is computed correctly, and the verifier checks that $CD^*=Z\cdot \Pi \mod q$, which means that $D^* = D\cdot \Pi \mod{q}$ is computed correctly. %and $\|Z\|_{\infty}\le \ell b$ (ensures that $\|D\|_{\infty}$ is small enough so that $Z=CD \mod q$ implies $Z=CD$ ). 

%Specifically, to prove that $\|D\|_\infty \le b$ for $D \in \mathbb{F}^{\ell \times m}$, the verifier picks a random binary matrix $C \gets \{0,1\}^{m \times \lambda}$ and sends it to the prover. The prover computes and sends $Z = DC$ to the verifier, who checks that $\|Z\|_\infty \le m b$. The verifier then randomly selects and sends a random vector $r \in \mathbb{F}_{q^t}^{\lambda \times 1}$, using the column linear relation proof protocol to verify that $Zr = D(Cr)$ is computed correctly.

The formalized description is presented in Figure \ref{fig.range1}:
\begin{figure}
	%\begin{algorithmic}[H]
    \algrule{1.2pt}
    \algrule{0.4pt}\smallskip  
		\normalsize
        \textbf{Range proof protocol:} For a commitment $\textsf{com}$ has been proven using \textsf{Testing} protocol, the prover wants to prove that the committed $D$ satisfying $\|D[i,\cdot]\|_2 \le b\sqrt{m}$ for each row $D[i,\cdot]$. Both parties run as follows:\smallskip\\        
        \begin{enumerate}
            \item The verifier chooses a random matrix $\Pi \gets \mathcal{D}^{m \times 256}$, where $\mathcal{D}$ is the distribution that $\Pr[0\gets \mathcal{D}]=0.5$ and $\Pr[-1\gets \mathcal{D}]=\Pr[1\gets \mathcal{D}]=0.25$, and sends $\Pi$ to the prover.
            \item The prover computes and sends $D^* = D\cdot \Pi \mod{q}$ to the verifier.
            \item The verifier checks that $\|D^*[i,\cdot]\|_2 \le b\sqrt{337m}$ for each $i\in [\ell]$. If so, send a random matrix $C \in \{0,1\}^{\lambda \times \ell}$ to the prover.
            \item The prover computes and sends $Z=CD \mod{q}$ to the verifier.
            \item Both parties run the $\textsf{Proving}$ protocol in Figure \ref{fig.linearproof} for the claim $Z=CD \mod{q}$. The verifier additionally checks $CD^*=Z\cdot \Pi \mod{q}$.
        \end{enumerate} 
    \algrule{0.4pt}
	%\end{algorithmic}
\caption{The First Range Proof Protocol}
	\label{fig.range1}
\end{figure}
\begin{lemma}
    Suppose that $q\ge 125\cdot b\cdot \sqrt{\frac{337m}{30}}$, then the protocol described in Figure~\ref{fig.range1} is an argument of knowledge for the statement that the committed $D$ satisfying that for each $i\in [\ell]$, $$\|D[i,\cdot]\|_2 \le b\cdot \sqrt{\frac{337m}{30}}.$$
\end{lemma}
\noindent\textit{Proof.} We first need to introduce two lemmas.

\begin{lemma}\label{JL1}\cite[Corollary 3.2]{GHL22} Let $\mathcal{D}$ be the distribution that $\Pr[0\gets \mathcal{D}]=0.5$ and $\Pr[-1\gets \mathcal{D}]=\Pr[1\gets \mathcal{D}]=0.25$, then for every vector $w\in \mathbb{Z}^d$ we have:
\begin{gather}
    \Pr_{\Pi \gets \mathcal{D}^{d \times 256}}[\|w\cdot\Pi\|_2<\sqrt{30}\|w\|_2]\lesssim 2^{-128}\\
    \Pr_{\Pi \gets \mathcal{D}^{d \times 256}}[\|w\cdot\Pi\|_2>\sqrt{337}\|w\|_2]\lesssim 2^{-128}
\end{gather}
\end{lemma}
Ward Beullens and Gregor Seile strengthen the first inequality so that it holds even modulo $q$:
\begin{lemma}\label{JL2}\cite[Lemma 4.2]{BS22} Let $q\in \mathbb{N}$ and let $\mathcal{D}$ be the distribution as above, then for every vector $w\in [\pm q/2]^d$ with $\|w\|_{2}\ge a$ for some bound $a\le q/125$, we have: 
   \begin{gather} \Pr_{\Pi \gets \mathcal{D}^{d \times 256}}[\|w\cdot\Pi\mod q\|_2<\sqrt{30}a]\lesssim 2^{-128}\end{gather}
\end{lemma}

Next, we first prove the \textbf{Completeness} of our protocol. When $\|D[i,\cdot]\|_2 \le b\sqrt{m}$, by Lemma~\ref{JL1}, we have $ \|D[i,\cdot] \cdot \Pi\|_2 \le \sqrt{337}\|D[i,\cdot]\|_2 = b\sqrt{337m}$, except with negligible probability. Since $q \ge 125 \cdot b \cdot \sqrt{\frac{337m}{30}} > b\sqrt{337m}$, it follows that $ D^*[i,\cdot]=D[i,\cdot]\cdot \Pi \bmod q= D[i,\cdot]\cdot \Pi$. Thereby we have that $\|D^*[i,\cdot]\|_2 < b\sqrt{337m}$, and completeness holds.

Finally, we prove the knowledge soundness of the scheme. Since the commitment passes the Testing protocol, we can use the knowledge extractor of that protocol to extract the committed database $D$. We now show that if $D$ does not satisfy $\|D[i,\cdot]\|_2 \le b \cdot \sqrt{\frac{337m}{30}}$, then the probability of passing the range proof is negligible.

When the prover returns the correctly computed $D^* = D \cdot \Pi \bmod q$ in Step 2, by Lemma~\ref{JL2} and the fact that $b \cdot \sqrt{\frac{337m}{30}} \le q/125$, if there exists an $i$ such that $\|D[i,\cdot]\|_2 > b \cdot \sqrt{\frac{337m}{30}}$, then for $w = D[i,\cdot]$, we have
\[
\Pr_{\Pi \gets \mathcal{D}^{d \times 256}}\bigl[ \|w\cdot\Pi \bmod q\|_2 < \sqrt{30}\,b \cdot \sqrt{\frac{337m}{30}} \bigr] \lesssim 2^{-128},
\]
which is negligible.

When the prover returns $D^* \neq D \cdot \Pi \bmod q$ in Step 2, consider the random matrix $C$ sent in Step 3. If the prover returns $Z = C D \bmod q$, then $C \cdot D \cdot \Pi \neq C \cdot D^* \pmod q$ except with negligible probability. If instead $Z \neq C D \bmod q$, then by the soundness of the Proving protocol, the probability that the adversary passes the verification in Step 5 is negligible.
\qed \medskip

\noindent\textit{Second approach.} This approach is based on range proof techniques (lookup techniques) commonly used in SNARKs. To prove that $\|D\|_\infty \le b$, we list all values in $[-b,b]$ as a table $\mathbf{t}=(-b,-b+1,\cdots,b)$. Then, we use lookup arguments to prove that all elements in $D$ lie in the table $\mathbf{t}$. 

%we first perform a binary decomposition of each entry of $D$, obtaining a new matrix $D_g$. We then prove that: (1) the binary decomposition is correct, and (2) every entry of $D_g$ is indeed either $0$ or $1$ (i.e., $x(1-x)=0$ for all entries $x$ of $D_g$).

To be more detailed, let $f_D$ be the multilinear extension of $D$, that is, $f_D$ is the polynomial such that $f_{D}(x,y)=D[\textsf{int}(x),\textsf{int}(y)]$ where $x\in \{0,1\}^{\log{\ell}}$, $y\in \{0,1\}^{\log m}$\footnote{We assume \(\ell\) and \(m\) are powers of two. Otherwise, we pad \(D\) with zeros at the bottom and right to obtain a suitable \(D'\), and perform the following proof on \(D'\). Notably, checking row linear relations on \(D'\) reduces to checking those on \(D\).}, and $\textsf{int}$ interprets each binary string as its corresponding integer. Then, following \cite{Habock22a,EFG22}, to prove that the values of $D$ lie in $\mathbf{t}$, we only need to prove that:
$$\sum_{\substack{(x,y)\in\{0,1\}^{\log\ell+\log m}}} \frac{1}{X+f_D(x,y)}=\sum_{x'\in [-b,b]}\frac{m_{x'}}{X+\mathbf{t}(x')}$$
where $m_{x'}$ means the number of $\mathbf{t}(x')=x'$ appears in $D$. Let the prover send  $\{m_{x'}\}$ to the verifier. 

%Write $\{m_{x'}\}$ as a $2^{b/2}\times 2^{b/2}$ matrix/database, and let the prover commits to it and runs our $\textsf{Testing}$ protocol. 

Then, the verifier samples and sends $t'$ random challenges $r_k\gets \mathbb{F}_{q}$, and the range check is reduced to $$\sum_{\substack{(x,y)\in\{0,1\}^{\log\ell+\log m}}} \frac{1}{r_k+f_D(x,y)}=\sum_{x'\in [-b,b]}\frac{m_{x'}}{r_k+\mathbf{t}(x')}$$ for all $k\in [t']$, where $t'$ is the integer such that $(\frac{\ell m+b}{q})^{^{t'}}\ge 2^{-\lambda}$. 

The verifier computes $\alpha_k=\sum_{x'\in [-b,b]}\frac{m_{x'}}{r_k+\mathbf{t}(x')}$ itself\footnote{When $b$ is too large, the computation of $\sum_{x'\in [b]}\frac{m_{x'}}{r_k+\mathbf{t}(x')}$ can be done with the help of prover. That is, the prover commits the multilinear extension polynomial of $m_{x'}$ using a polynomial commitment, rather than directly sending all $m_{x'}$. After receiving $r_k$, the prover computes and sends all $\alpha_k$, and then proves their correctness using sumchecks in a similar way as how it proves $\alpha_k=\sum_{\substack{(x,y)\in\{0,1\}^{\log\ell+\log m}}} \frac{1}{r_k+f_D(x,y)}$. }. 
The prover computes the databases denoted by $D_k[x,y]=(r_k+D[x,y])^{-1}$, commits them and sends the commitments to the verifier. Both parties runs our $\textsf{Testing}$ protocols for these commitments. Now, let $f_{D_k}$ be the corresponding multilinear polynomials, just like $f_D$. The verifier samples and sends randomness $(r_x,r_y)\in \mathbb{F}_{q^t}^{{\log\ell+\log m}}$ to the prover, and the statements is now reduced to the following sumchecks, for all $k$:
\begin{align}
    &\sum_{\substack{(x,y)\in\{0,1\}^{\log\ell+\log m}}} f_{D_k}(x,y)=\alpha_k\\
    %\sum_{\substack{(x,y)\in\{0,1\}^{\log\ell+\log m}}} eq((x',y'),(x,y))\cdot (f_{D_k}(x,y)\cdot (r_k+f_D(x,y))-1)=0\\
    \begin{split}
&\sum_{\substack{(x,y)\in\{0,1\}^{\log\ell+\log m}}} 
  \mathrm{eq}\bigl((r_x,r_y),(x,y)\bigr) \\
&\qquad \cdot \Bigl( f_{D_k}(x,y)\cdot \bigl(r_k+f_D(x,y)\bigr)-1 \Bigr) = 0
\end{split}
\end{align}

Both parties run the classic sumcheck protocols for them (batched using random linear combinations for better efficiency), which reduces the checks to verifying $f_D(x',y') = t$ for some $x',y'$ (and $f_{D_k}(x'_k, y'_k) = t_k$ for some $x'_k, y'_k$). This is equivalent to proving a relation of the form $x^* \cdot D \cdot (y^*)^\perp = t$, where
\[
x^* = \bigotimes_{i=0}^{a-1} (x'[i], 1-x'[i]), \qquad y^* = \bigotimes_{i=0}^{b-1} (y'[i], 1-y'[i]).
\]
The prover sends $x^* \cdot D$ to the verifier, and the two parties run a row linear relation proof to verify the correctness of this computation. Finally, the verifier computes $(x^{*} \cdot D) \cdot (y^{*})^\perp$ and checks that it equals $t$ (the checks for $f_{D_k}(x'_k, y'_k) = t_k$ can be done in the same method).

A complete description of the second range proof and its security proof are deferred to Appendix~\ref{APP:rangeproof2}.

\begin{remark}[Fiat-Shamir Heuristic \cite{36664.36676}] Since all the protocols described above are public-coin arguments of knowledge, they can be transformed into non-interactive ones using the Fiat--Shamir heuristic. Roughly, the Fiat--Shamir heuristic replaces all random verifier challenges with the output of a cryptographic hash function applied to the transcript so far. The resulting non-interactive proof consists solely of the prover's messages, and anyone can verify it without further interaction. This transformation preserves the (knowledge) soundness in the random oracle model.
\end{remark}

\section{Verifiable PIR with efficient online phase}
The commitment, linear relation proof, and range proof schemes let us build
Verifiable PIR without the SIS assumption.  Our construction builds on VeriSimplePIR, and the description is as follows:

During the one-time setup phase, for the public matrix \(A\) used by the Regev scheme and database $D\in \mathbb{Z}_p^{\ell\times m}$, the server computes \(H = DA\mod q\) as the hint for following Regev's encryption, where $q$ is its ciphertext module, $p$ is its plaintext module, while committing to \(D\) in $\mathbb{F}_{q_{\text{com}}}$ as the digest, where $\mathbb{F}_{q_{\text{com}}}$ is the field using by our commitment and corresponding proof protocols. It also runs the testing and the range-checking protocols to verify the well-formedness of the commitment and the norm of \(D\). 

In the proof preprocessing phase, the client selects a random matrix $C\gets \{0,1\}^{\kappa\times \ell}$ and sends \(\text{Enc}(C)\) to the server (using a FHE scheme with a plaintext module $q_{\text{FHE}}$). The  server homomorphically computes and returns \(\text{Enc}(C D)\). The client decrypts to recover \(Z=C D\). Both parties then use the commitment to \(D\) to execute a (batched) proof for the linear combination \(Z=CD\) (here we require that $q_\text{com}$ and $q_{\text{FHE}}$ be large enough so that $CD=CD\bmod{q_\text{com}}=CD\bmod{q_{\text{FHE}}}$, so that $C$ and $Z$ can be used to check the column linear relation of $D$ over any ring).  The verifier then checks $CH=ZA\bmod{q}$ to ensure that $H=DA\bmod{q}$. 

In the online phase, for each query, the client computes and sends \(u=As+e+\textsf{Ecd}(\mu)\bmod{q}\) to the server. The server returns \(v = Du\bmod{q}\). The client checks whether \(Z \cdot u = C\cdot v\bmod{q}\) holds. If the check passes, the client decrypts to obtain \(\textsf{Dcd}(v-Hs)\).

The formalized construction is shown in Figure~\ref{fig.vPIR}.

\begin{figure}
\algrule{1.2pt}
\algrule{0.4pt}\smallskip  
	%\begin{algorithmic}[H]
     %\begin{flushleft}
		\normalsize
        \textbf{vPIR Protocol:} 
        Let \(A \in \mathbb{Z}_q^{m \times n}\) be a random public matrix used for Regev encryption, where \(q\) may be a power of \(2\). The server holds a private matrix \(D \in \mathbb{Z}_p^{\ell \times m}\) ($D[i,j]\in (-p/2,p/2]$). Let \(q_{\text{com}},q_\text{FHE}\) be the primes used by the commitment scheme and FHE scheme's plaintext modulus. 
The client holds a secret key for the FHE scheme. The protocol then proceeds as follows:\smallskip\\
        %Let $A\in \mathbb{Z}^{m\times n}_q$ be a random public matrix used for Regev encryption, {where $q$ can be a power of 2}. The server holds a private matrix $D\in \mathbb{Z}^{\ell\times m}$ satisfying $\|D\|_{\infty}\le p$. Let $q''$ be a prime such that $q''\ge \ell p$, and $q'$ be a prime such that: when using first range proof, $q'\ge \ell p\sqrt{\frac{337m}{30}}$; when using the second range proof, $q'\ge \ell p$. The client holds a secret key for the BGV/BFV scheme with plaintext module $q''$. The protocol proceeds as follows:\smallskip\\
        \begin{enumerate}
            \item \textbf{One-time setup phase} (database-level preprocessing).
            \begin{enumerate}
                \item The server computes $H = DA \mod{q}$ and generates a commitment $\textsf{com}$ to $D$ over $\mathbb{F}_{q_{\text{com}}}$.
                \item The server runs non-interactive versions (obtained via the Fiat--Shamir heuristic) of the testing protocol and the range-proof (Figure~\ref{fig.range1} or~\ref{fig.range2}) to obtain the proofs $\pi_1,\pi_2$ for the commitment's well-formedness and the norm of $D$ (i.e., $\|D\|_\infty \le p/2$ or $\|D\|_2 \le \sqrt{m}\,p/2$).
                \item The server outputs $H,\textsf{com},\pi_1,\pi_2$ as the digest. And the client checks the validity of $\pi_1$ and $\pi_2$. 
                %The server runs a testing protocol and a range-proof (Figure \ref{fig.range1} or \ref{fig.range2}) to prove the commitment's well-formedness and the norm of $D$ (i.e., $\|D\|_\infty \le p$ or $\|D\|_2\le \sqrt{m}p$). 
            \end{enumerate}
            
            \item \textbf{Proof preprocessing phase} (client-level preprocessing).
            \begin{enumerate}
                \item The client selects a random matrix $C \gets \{0,1\}^{\kappa\times \ell}$, where $\kappa$ is the statistic security parameter, and sends $\text{Enc}(C)$ to the server using the FHE encryption scheme.
                \item The server computes and sends $\text{Enc}(C\cdot D)$.
                \item The client decrypts to obtain $Z = C \cdot D\bmod{q_{\text{FHE}}}$, and checks $C\cdot H=Z\cdot A \bmod{q}$.
                \item Using the commitment to $D$, both parties execute a batched linear relation proof (Figure~\ref{fig.linearproof}) to prove the linear relation $Z = CD \bmod q_{\text{com}}$.
            \end{enumerate}
            
            \item \textbf{Online phase (per query).}
            \begin{enumerate}
                \item The client computes and sends $u = A s + e + \textsf{Ecd}(\mu)\mod{q}$ to the server, where $s$ is a random secret, $e$ is a small error, and $\mu$ is the query vector.
                \item The server returns $v = D u\mod{q}$ to the client.
                \item The client checks whether $Z \cdot u = C \cdot v \mod{q}$.
                If all checks pass, the client proceeds; otherwise, it aborts.
                \item The client computes $\textsf{Dcd}(v - H s)$, which recovers the target message $D\cdot\mu$.
            \end{enumerate}
        \end{enumerate} 
    % \end{flushleft}
	%\end{algorithmic}
\algrule{0.4pt}
\caption{Formal description of the Verifiable PIR protocol}
	\label{fig.vPIR}
\end{figure}\medskip

\noindent\textbf{Parameter Selection.} To ensure the completeness and security of our construction, we require that \(q_{\text{FHE}} \ge \ell p\); moreover, if the first range proof is used, then \(q_{\text{com}} \ge \ell p \sqrt{\frac{337m}{30}}\) and \(q \ge p^2 \cdot \sigma \cdot \sqrt{\frac{337m \ln(2/\delta)}{15}}\), while if the second range proof is used, then \(q_{\text{com}} \ge \ell p\) and \(q \ge p^2 \cdot \sigma \cdot \sqrt{2m \ln(2/\delta)}\).

\begin{theorem}The scheme described in Figure~\ref{fig.vPIR} is a verifiable PIR scheme if the parameter selection satisfies the above requirements and $(n, m, q, \chi)$-LWE problem is hard. 
\end{theorem}
\begin{proof} \textbf{Completeness} follows from the underlying protocol's completeness and the fact that, for any database $D\in \mathbb{Z}_p^{\ell\times m}$, $C\in \{0,1\}^{\kappa\times\ell}$, and primes $q_{\text{com}},q_{\text{FHE}}$ larger than $\ell p$, $CD=CD\bmod{q_{\text{com}}}=CD\bmod{q_{\text{FHE}}}$.

The \textbf{Digest Binding} property directly follows from the binding property of our commitment scheme and the knowledge soundness property of underlying $\textsf{Testing}$ protocol. 

Now we prove the \textbf{Soundness} as follows. First we introduce the following statistical lemma:
\begin{lemma}\label{guassl}\cite[Lemma 2.4]{CastroL24} Let $\ell,\kappa\in \mathbb{N}$ and $p\ge 2$. For all $x\in \mathbb{Z}^\ell_p$ such that $x\neq 0$, the following holds:
$$\Pr_{C\gets \{0,1\}^{\kappa\times\ell}}[C\cdot x=0]\le 2^{-\kappa}.$$ 
\end{lemma}
Let $D$ be the database extracted from the digest via the digest binding property. First, from the soundness of our range proof protocol, when the proof is acceptable, we have that 1) $\|D\|_2<  \sqrt{\frac{337m}{30}}\cdot p$ when the first range-proof is used, or 2) $\|D\|_{\infty}< p$ when the second range-proof is used. 

Next, we show that the probability that an adversary succeeds in making the client accept an incorrect result over a polynomial number of queries is negligible. Conversely, if the adversary succeeds on the $k$-th query, then at least one of the following events must occur:
\begin{itemize}
    \item Event 1: $Z \neq CD \pmod q$ or $H \neq DA \pmod q$.
    \item Event 2: Event 1 does not hold, and the client's decryption result is not equal to $D \cdot \mu_k$
\end{itemize}
 
First, we prove that the probability of the client accepts $k$-th query but Event 1 occurring is negligible. By the soundness of our linear relation proof, we have \(Z = CD \pmod{q_{\text{com}}}\) except with negligible probability. Since \(C\) is a binary matrix, and \(q_\text{com} \ge \ell \|D\|_{\infty}\), it follows that \(CD \bmod q_\text{com} = CD \bmod q\). Hence \(Z = CD \pmod{q}\). Furthermore, by Lemma~\ref{guassl}, the probability that \(H \neq DA \pmod{q}\) but \(CH = C(DA) \pmod{q}\) is negligible.

Second, we prove that the probability of the client accepts $k$-th query, $Z = CD \pmod q, H = DA \pmod q$ but Event 2 occurring occurring is negligible. We directly construct a reduction algorithm \(\mathcal{D}^{\mathcal{A}}\) such that if above event occurs with noticeable probability, then \(\mathcal{D}^{\mathcal{A}}\) can break the IND-CPA security of the FHE scheme. Specifically, \(\mathcal{D}^{\mathcal{A}}\) is constructed as follows:

\begin{enumerate}
    \item It acts as the client and interacts with \(\mathcal{A}\) until the end of the one-time setup phase. If the digest verification fails, it aborts; otherwise, it uses the extractor to obtain the committed database \(D\) from \(\mathcal{A}\).
    \item It picks two random matrices \(C_0, C_1\), sends them to the IND-CPA challenger, and receives a challenge ciphertext \(\text{Enc}(C_b)\). It forwards this ciphertext to \(\mathcal{A}\) and, after receiving a response, proceeds to the query phase.
    \item It samples a random query $k^*$, and then repeats the following query phase until the $k^*$-th query: \(\mathcal{D}^{\mathcal{A}}\) sends honestly generated \(u\) to the server. When the server returns \(v\), it checks whether \(v = Du\). If so, it return accept to $\mathcal{A}$; otherwise, it returns \(\bot\) (reject) to \(\mathcal{A}\) and proceeds to the next query.
    \item In the $k^*$-th query, \(\mathcal{D}^{\mathcal{A}}\) sends an honestly generated \(u_k\) to the server. When the server returns \(v_k\), it checks whether \(v_k = Du_k\). If so, output a random $b^*\gets \{0,1\}$. Otherwise, it checks whether there exists a \(b^*\) such that \(C_{b^*}\cdot v_k = C_{b^*}\cdot D u_k\). If such a \(b^{*}\) exists, outputs \(b^*\); otherwise, output a random $b^*\gets \{0,1\}$. 
\end{enumerate}

We complete the proof in two steps:
\begin{enumerate}
    \item The probability that \(C_{b^*}\cdot v = C_{b^*}\cdot D u\) occurs in step 4 while \(b \neq b^*\) is negligible. 
    Observe that the adversary receives \textit{no information} about \(C_{1-b}\) (including its encryption), which makes \(C_{1-b}\) independent of \(u\) and \(v\). Therefore, by Lemma~\ref{guassl}, the probability that \(C_{1-b}\cdot v = C_{1-b}\cdot D u\) holds is negligible.

    \item The probability that \(C_{b^*}\cdot v = C_{b^*}\cdot D u\) occurs in step 4 is noticeable.
    By assumption, there exists a \(k\) s.t., the event ``\textit{the client accepts the $k$-th query, \(Z = CD \pmod q\), \(H = DA \pmod q\) but the client's decryption result is not equal to \(D \cdot \mu_k\)}'' happens with a noticeable probability. {From the range proof and our parameter choices, we have that if the client's decryption result is not equal to \(D \cdot \mu_k\), then necessarily either \(H \neq DA \pmod q\) or \(v \neq Du \pmod q\) (see section 3.1). }Therefore, we simplify the above event to the following: there exists a \(k\) s.t., the probability that \(v_k \neq Du_k \pmod q\) yet \(C \cdot v_k = C \cdot D u_k\) occurs for the first time is noticeable. When \(k^* = k\), this event coincides exactly with the event that \(C_{b^*} \cdot v = C_{b^*} \cdot D u\) occurs during the execution of \(\mathcal{D}^{\mathcal{A}}\). Since the total number of queries is polynomial, the probability that \(C_{b^*} \cdot v = C_{b^*} \cdot D u\) occurs is noticeable. This completes the proof of soundness.
\end{enumerate}\medskip

\noindent\textbf{Query Hiding}. This property can be proved simply and directly by constructing a simulator as follows. The simulator acts as an honest client and interacts with the server until the end of the preprocessing phase. If the proofs and checks in the preprocessing phase are accepted, it proceeds to the query phase; otherwise, it aborts the protocol. During a polynomial number of queries, the simulator always chooses a random \(u\) and sends it to the server. When the server returns \(v\), the simulator checks whether \(Zu=Cv\) and returns the result of this check. Finally, the simulator outputs the adversary's view.

Since the only difference between the simulator and a real client is that the simulator sends a random \(u\) while the client sends a Regev encryption \(u = A s + \Delta \mu + e\), by the LWE assumption the two distributions are computationally indistinguishable.
\end{proof}

\begin{remark}[{When $q$ is a large prime}]
In this case, we can simply set $q = q' = q''$. Moreover, the matrix $C$ no longer needs to be chosen from $\{0,1\}^{\kappa \times \ell}$; instead, it can be chosen from $\mathbb{F}_q^{t \times \ell}$, where $t$ is the smallest integer such that $p^t > 2^{128}$, thereby reducing the offline communication. 
\end{remark}
{\begin{remark}[Assumptions] Our verifiable PIR construction instantiated with the second range proof relies on LWE, collision-resistant hashing, and the random-oracle model, whereas the construction instantiated with the first range proof additionally relies on the heuristic assumption underlying Lemmas~\ref{JL1} and~\ref{JL2}. In contrast, VeriSimplePIR~\cite{CastroL24} relies on LWE, SIS, and the random-oracle model.
\end{remark}}

\section{Performance}

\subsection{Complexity Analysis}
%由于方案大部分计算都位于

%one-time setup阶段：与verisimple PIR相比，除了H=DA的计算（需要约$\ell m n$环/域乘法），我们需要额外计算承诺和相关的证明。在使用精确的范围检查时，仅需要约O（\ell m \logm）域乘法；在使用almost的范围检查时，仅需要约O（\ell m \logm）域乘法和 O（\ell m 256）环加法。从复杂度上看，我们与verisimple PIR相比并没有复杂度的提高（remind that n=1024~2024）

\noindent The \textbf{Offline-phase} complexity analysis is as follows.\smallskip

{\noindent\textit{Offline server computation.} The server's computation splits into two parts: digest generation and client-level preprocessing. In the first part, VeriSimplePIR mainly requires computing \(H = DA\) (roughly \(O(\ell m n)\) multiplications), whereas our server additionally computes our commitment and the associated proofs. With the first range proof, it adds about \(O(\ell m \log m)\) field multiplications and \(O(\lambda \ell m)\) ring additions. With the second range proof, this adds only about \(O(\ell m \log m)\) field multiplications. 

In the second part, VeriSimplePIR mainly requires homomorphically computing \(Z = CD\) (at least \(O(\ell m \lambda)\) multiplications), whereas our scheme additionally computes the proof for \(Z = CD\), this only requires the server to send some hash paths and incurs no extra computation. 

Therefore, from a complexity perspective, our scheme incurs no asymptotic overhead over VeriSimplePIR in either part (recall that \(n\) is typically set between \(1024\) and \(2048\)). \smallskip}

%Compared to VeriSimplePIR, which mainly requires computing \(H = DA\) and homomorphically compute $Z=CD$ (roughly \(O(\ell m n)\) ring/field multiplications),  servers in our scheme additionally computes the commitment and the associated proofs. With the second range check proof, this adds only about \(O(\ell m \log m)\) field multiplications. With the first range check proof, it adds about \(O(\ell m \log m)\) field multiplications and \(O(\lambda\cdot\ell\cdot m)\) ring additions. From a complexity perspective, our scheme does not incur any asymptotic overhead over VeriSimplePIR (recall that $n$ is typically set between $1024$ and $2048$).\smallskip

\noindent\textit{Offline client computation.} The client verifies the digest and runs the proof preprocessing phase. Compared to VeriSimplePIR, which mainly requires checking \(CH = ZA\) and encrypting \(C\) as well as decrypting \(\textsf{Enc}(Z)\) (roughly \(O(\kappa \cdot (\ell + m) \cdot n)\) ring/field multiplications), our scheme additionally requires the client to verify the underlying linear relation proofs and range‑check proofs. Depending on the range proof used, this adds about \(O(\lambda \ell + m \log m + p)\) or \(O(\lambda \ell + m \log m + \lambda \cdot \kappa \cdot (\ell + m))\) ring/field multiplications. (When \(p\) is large, as previously discussed, this computation can be offloaded to the server.) From a complexity perspective, our scheme does not incur any asymptotic overhead over VeriSimplePIR.\smallskip

\noindent\textit{Offline communication.} As in VeriSimplePIR, the main communication overhead of our scheme comes from sending \(H = DA\), which requires \(O(\ell n)\) elements. The remaining components, amounting to \(O(\lambda (\ell + m) + p)\), are comparatively smaller. (When \(p\) is large, as previously discussed, this communication can be offloaded to the server.) From a complexity perspective, our scheme does not incur any asymptotic overhead over VeriSimplePIR.\medskip

\noindent The \textbf{Online-phase} complexity analysis is as follows.

 As in VeriSimplePIR, the communication overhead of our scheme in the query phase is only \((\ell + m)\) field/ring elements. The client’s computation requires \(\ell n + \kappa(\ell + m)\) field/ring operations, while the server’s computation involves \(\ell \cdot m\) field/ring elements. However, due to differences in parameter selection, for the same other parameters, our scheme allows a larger plaintext modulus \(p\). Consequently, for a database of the same size, we can pack more messages together, resulting in a packed matrix with smaller \(\ell\) and \(m\). For an 8 GB database, VeriSimplePIR uses \(p = 2^{16}\), whereas our scheme can achieve \(p = 2^{24}\) or \(2^{25}\) (depending on the range proof used). In the case of 1‑bit data, this \textit{theoretically} reduces the communication overhead and client computation time by about 20\%, and reduces the server computation time by over 35\%.

\subsection{Implementation}

We implement our protocol in C++, building over the codebase of
 VeriSimplePIR \cite{CastroL24}.  For the {proof preprocessing phase}, we instantiate the
{homomorphic matrix multiplication} $Z = C \cdot D$ using the same
{ LHE}~\cite{CastroL24} as VeriSimplePIR's preprocessing,
performing row-wise encryption of the challenge matrix $C$.
For the {online phase}, we reuse the
LHE scheme from SimplePIR with the
PackedMatrix column-packing optimization. \smallskip%We provide our online-phase artifact through Anonymous GitHub\footnote{Anonymous artifact:\url{https://anonymous.4open.science/r/vpir_final-AC50}}.

%\noindent\textbf{Optimizations and assumptions.}
%We \textcolor{red}{eliminate the honest digest assumption}
%required in prior verifiable PIR protocols~\cite{CastroL24}:
%our \textcolor{red}{pre-proof} phase verifies
%$\textcolor{red}{CH = ZA}$ over $\mathbb{Z}_{2^{64}}$,
%cryptographically binding $H$ to the committed $D$.We adopt the \textcolor{red}{database-level preprocessing}
%optimization~\cite{10.5555/3620237.3620455} and select
%\textcolor{red}{LWE parameters} under the \textcolor{red}{L2-norm correctness bound}
%($\|D\|_2 \leq 2^d\sqrt{m}$ from our relaxed range proof),
%matching SimplePIR's tight parameter regime without the
%$2\ell p$ slack of VeriSimplePIR's SIS-based analysis. In summary, we can achieve the security of VerifiablePIR by using the parameters of SimplePIR, and also obtain satisfactory experimental results.

\noindent\textbf{Parameters.}
  %All database entries are split into
  %$d = 8$-bit chunks (1 byte per entry).
  For the online LHE, we set the LWE dimension
  $n = 2048$ and use ciphertext modulus
  $q = 2^{64}$ (implicit in the
  \texttt{uint64\_t} datatype) as same as VeriSimplePIR, achieving 110-bit security estimated by Lattice Estimator\footnote{Lattice security levels were estimated using the
\textit{Lattice Estimator}
(\url{https://github.com/malb/lattice-estimator},
commit \texttt{5ba00f5}) with the MATZOV cost model
(\texttt{estimator.reduction.MATZOV}).}. The plaintext modulus
  $p$ chosen dynamically
  (typically $p \in [2^8, 2^{24}]$) to minimize per-query
  communication.
  We sample error from a {discrete Gaussian
  distribution} with standard deviation
  $\sigma = 6.4$ \cite{dgs}, and sample the
  {LWE secret key} uniformly over
  $\mathbb{Z}_q$.
  For the {proof preprocessing phase}, %we use a
  %\textcolor{red}{Multi-Limb LHE}~\cite{CastroL24} with the same
  %LWE dimension $n = 2048$, plaintext modulus
  %$\textcolor{red}{p \cdot \ell}$ (to avoid overflow in
  %homomorphic accumulation), and ciphertext space
  %$\textcolor{red}{Q = 2^{64} \times \kappa'}$ where
  %$\textcolor{red}{\kappa' = \lceil\sqrt{\ell m}\,\rceil}$
  %(rounded to odd).
  the {commitment} operates over the
  {64-bit Solinas prime}
  ${p_{\text{com}} = 2^{64} - 2^{32} + 1}$
  and its quadratic extension
${\mathbb{F}_{p_{\text{com}}^2}}$ \cite{10.1145/3719027.3765226},
  with RS code rate $1/2$ and proximity parameter
  ${\beta = 0.25}$.
  The {tight range proof} uses
  ${\mathrm{t'} = 3}$ sumcheck challenges.

\iffalse
  \noindent\textbf{Packed storage and computation.}
 Our implementation employs \textcolor{red}{three levels} of
  packing to reduce memory and accelerate online operations.
  First, \textcolor{red}{$d = 8$}-bit database entries are packed
  into \textcolor{red}{$\mathbb{Z}_p$} elements,
  with $\lfloor\log p / d\rfloor$ entries per element.
  The matrix $D$ is then stored in
  \textcolor{red}{column-packed} format
  (\textcolor{red}{PackedMatrix}), where each
  \textcolor{red}{64-bit} word holds up to
  $\lfloor 64 / \log p \rfloor$ entries from consecutive columns
  of the same row, halving memory footprint.
  The \textcolor{red}{Answer} routine ($v = D \cdot u$) uses a
  \textcolor{red}{hand-unrolled} column-major kernel processing
  \textcolor{red}{8 rows simultaneously} with independent
  accumulators, extracting packed entries via bit-shift and mask.
  The {Prove} and {proof preprocessing}
  computations also operate directly on packed storage.
  For the {range proofs}, packed entries are
  unpacked back to individual $d$-bit values to certify norm bounds on
  the original database entries.
  \fi

\subsection{Performance Analysis}    

\noindent\textbf{Experiments Setup.}
All the experiments are run on AMD EPYC 9474F with two 48-Core Processor @3.6 GHz and 1.5 TB memory, and constrained in a single-thread. {Table~\ref{tab:performance_comparison} and Table~\ref{tab:online} report online
communication and latency: the former for databases ranging from
\(512\)\,MiB to \(16\)\,GiB with \(1\)-byte entries, the latter for
\(1\)\,GiB databases with entry sizes from \(1\)~bit to \(32\)~bytes.
%Table~\ref{tab:client-prep} reports the offline client-level preprocessing costs\footnote{More precisely, the entire offline phase costs except the server-side digest generation, and thus also includes the client's digest verification and digest download.}. 
In these tables, Ours\(_1\) denotes our verifiable PIR scheme instantiated with the first range proof protocol, and Ours\(_2\) the one instantiated with the second.}
%Each benchmark instance was executed 100 times.

\begin{table*}[h]
\centering
\caption{Comparison of our scheme with VeriSimplePIR and SimplePIR.}
\label{tab:performance_comparison}
\begin{tabular}{l | ccc | cccc | cccc}
\toprule
\multirow{2}{*}{\textbf{DB Size}} 
    & \multicolumn{3}{c|}{\textbf{Offline Comm. (MiB)}} 
    & \multicolumn{4}{c|}{\textbf{Online Comm. (KiB)}} 
    & \multicolumn{4}{c}{\textbf{Online Latency (ms)}} \\
\cmidrule(lr){2-4} \cmidrule(lr){5-8} \cmidrule(l){9-12}
  (1-byte entries)  & \textbf{VeriSim.} & \textbf{Ours}\(_1\) & \textbf{Ours}\(_2\) 
    & \textbf{SimplePIR} & \textbf{VeriSim.} & \textbf{Ours}\(_1\) & \textbf{Ours}\(_2\) 
    & \textbf{SimplePIR} & \textbf{VeriSim.} & \textbf{Ours}\(_1\) & \textbf{Ours}\(_2\) \\
\midrule
512 MiB  & 793  & 773 & 978  & 209  & 256  & 209  & 209  & 53.9   & 81.5   & 57.1   & 56.6   \\
1 GiB    & 1121 & 1093 & 1355 & 296  & 362  & 296  & 296  & 98.5  & 139.1  & 101.9  & 101.6  \\
4 GiB    & 2243 & 2187 & 2583 & 592  & 724  & 592  & 592  & 322.1  & 474.5  & 328.8  & 329.2 \\
8 GiB    & 3173 & 3092 & 3600 & 836  & 1024 & 836  & 836  & 603.4  & 944.5 & 615.4  & 611.7  \\
16 GiB   & 4487 & 4374    & 5038 & 1182 & 1448 & 1182 & 1182 & 1161.4 & 1744.2 & 1176.3 & 1176.6 \\
\bottomrule
\end{tabular}
\end{table*}

\iffalse
\begin{table*}[t]
\color{red}
\centering
\caption{Computation and communication costs of client-level
preprocessing, for \(1\)-byte entries.}
\label{tab:client-prep}
\footnotesize
\begin{tabular}{@{}lcccc@{}}
\toprule
DB Size & SimplePIR (hint \(H\)) & VeriSimplePIR & Ours\(_1\) & Ours\(_2\) \\
\midrule
256\,MB & (\(-\), 148\,MB)   & (12.164\,s, 560\,MB)  & (12.181\,s, 546\,MB)  & (14.595\,s, 741\,MB) \\
1\,GB   & (\(-\), 296\,MB)   & (36.478\,s, 1116\,MB) & (33.184\,s, 1093\,MB) & (36.869\,s, 1337\,MB) \\
4\,GB   & (\(-\), 591\,MB)   & (118.431\,s, 2224\,MB)& (114.641\,s, 2186\,MB)& (123.149\,s, 2662\,MB) \\
8\,GB   & (\(-\), 836\,MB)   & (240.813\,s, 3149\,MB)& (233.742\,s, 3092\,MB)& (239.064\,s, 3766\,MB) \\
\bottomrule
\end{tabular}
\end{table*}
\fi

\begin{table}[t]
\centering
\caption{Online cost for a \(1\)~GB database. Each cell reports
communication (KiB)\,/\,latency (ms).}
\label{tab:online}
\footnotesize
\setlength{\tabcolsep}{3.5pt}
\begin{tabular}{@{}lcccc@{}}
\toprule
Entry Size & SimplePIR & VeriSimplePIR & Ours\(_1\) & Ours\(_2\) \\
\midrule
1\,bit  & (289.6/93.9) & (351.2/134.9) & (295.6/101.3) & (289.6/96.9) \\
2\,bit  & (295.6/99.1) & (362.0/139.6) & (295.6/101.4) & (295.6/102.3) \\
3\,bit  & (295.6/98.8) & (373.9/147.3) & (295.6/101.3) & (295.6/101.9) \\
4\,bit  & (295.6/98.6) & (362.0/135.7) & (295.6/101.5) & (295.6/101.9) \\
5\,bit  & (289.6/93.3) & (373.9/147.2) & (323.8/115.2) & (289.6/95.8) \\
6\,bit  & (295.6/98.8) & (418.0/178.3) & (295.6/101.5) & (295.6/101.8) \\
7\,bit  & (316.0/108.3) & (387.0/157.9) & (316.0/111.0) & (316.0/110.7) \\
1\,B    & (295.6/99.0) & (362.0/139.3) & (295.6/101.5) & (295.6/102.0) \\
2\,B    & (362.0/135.6) & (362.0/139.1) & (362.0/138.7) & (362.0/138.9) \\
4\,B    & (362.0/135.6) & (362.0/139.5) & (362.0/139.2) & (362.0/139.2) \\
8\,B    & (313.5/105.6) & (362.0/138.1) & (313.5/113.4) & (313.5/109.4) \\
16\,B   & (313.5/106.0) & (362.0/138.0) & (313.5/111.9) & (315.5/109.2) \\
32\,B   & (300.1/98.5) & (362.0/138.6) & (300.1/103.2) & (300.1/100.8) \\
\bottomrule
\end{tabular}
\end{table}

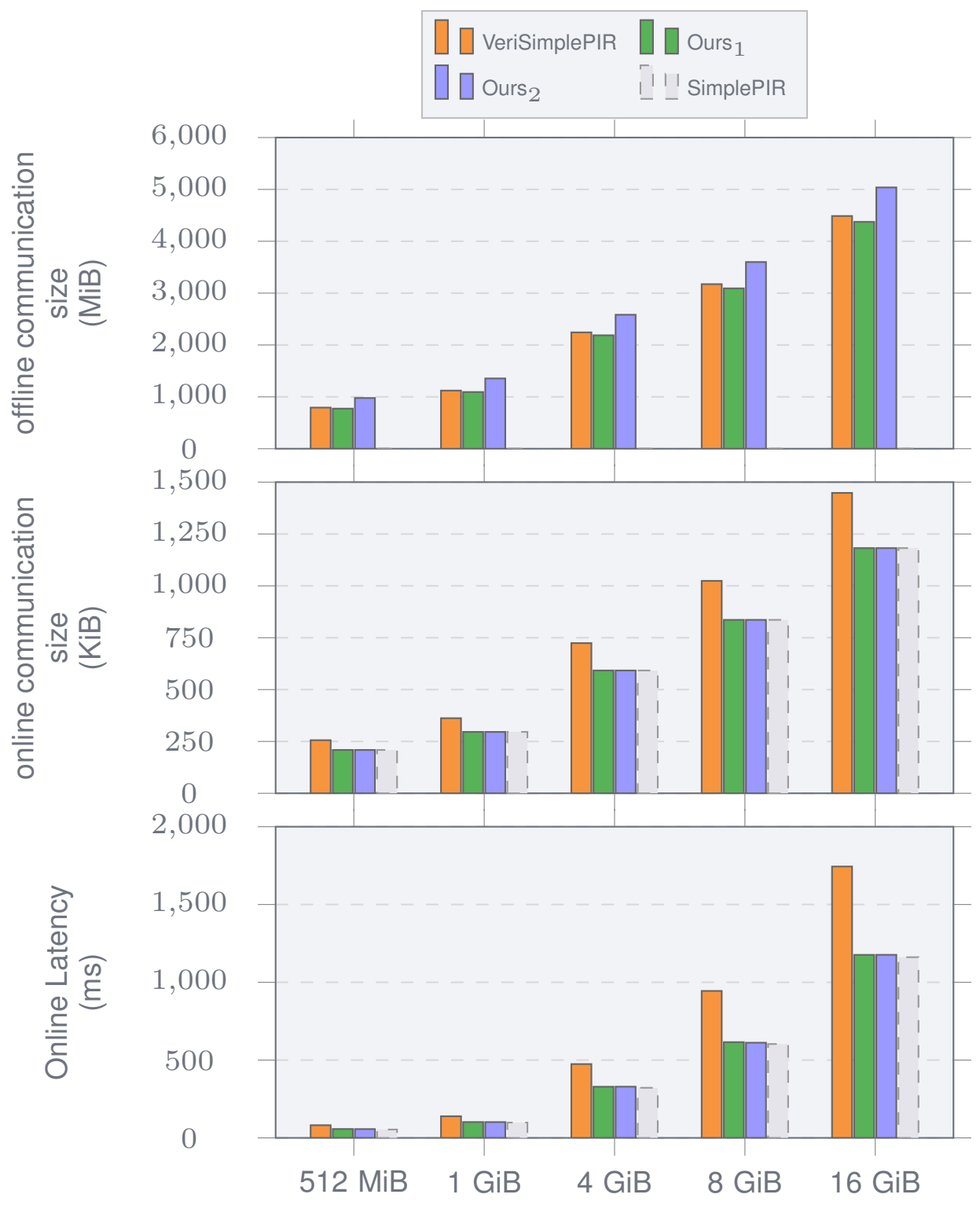
\begin{figure}[t]
    \centering

\begin{tikzpicture}

\pgfplotsset{
    every axis/.append style={
        ybar=0.5pt,
        bar width=4.8pt,
    }
}

\begin{groupplot}[
    group style={
        group size=1 by 3,
        vertical sep=8pt,
        x descriptions at=edge bottom,
    },
    width=0.85\columnwidth,
    height=4.2cm,
    enlarge x limits=0.15,
    symbolic x coords={
        512 MiB,
        1 GiB,
        4 GiB,
        8 GiB,
        16 GiB
    },
    xtick=data,
    axis background/.style={fill=color_bg},
    ymajorgrids=true,
    grid style={dashed,draw=gray!30},
    tick align=outside,
    axis line style={draw=color_text},
    x tick label style={
        text=color_text,
        font=\sffamily\scriptsize,
        align=center,
        text width=1cm,
        /pgf/number format/assume math mode=true,
    },
    y tick label style={
        text=color_text,
        align=center,
        font=\sffamily\scriptsize,
        text width=1cm,
        inner sep=2pt,
    },
    ylabel style={
        text=color_text,
        align=center,
        font=\sffamily\scriptsize
    },
]

% ============================================================
% 1. Offline communication
% 没有 SimplePIR 的 offline communication，因此只有三根柱子
% ============================================================

\nextgroupplot[
    ymin=0,
    ymax=6000,
    ytick={0,1000,2000,3000,4000,5000,6000},
    ylabel={offline communication\\size\\(MiB)},
    legend style={
        at={(0.5,1.06)},
        anchor=south,
        legend columns=2,
        legend cell align=left,
        draw=gray!50,
        fill=color_bg,
        text=color_text,
        font=\sffamily\tiny,
        /tikz/every even column/.append style={
            column sep=0.12cm
        }
    }
]

% SimplePIR 只加入图例，不绘制离线通信柱子
%\addlegendimage{
%    area legend,
%    fill=color_gray,
%    draw=gray!80,
%    dashed
%}

% VeriSimplePIR
\addplot[
    fill=color_orange,
    draw=black!60
] coordinates {
    (512 MiB,793)
    (1 GiB,1121)
    (4 GiB,2243)
    (8 GiB,3173)
    (16 GiB,4487)
};
\addlegendentry{VeriSimplePIR}

% Ours_1: Relax
\addplot[
    fill=color_green,
    draw=black!60
] coordinates {
    (512 MiB,773)
    (1 GiB,1093)
    (4 GiB,2187)
    (8 GiB,3092)
    (16 GiB,4374)
};
\addlegendentry{Ours\(_1\)}

% Ours_2: Tight
\addplot[
    fill=blue!40,
    draw=black!60
] coordinates {
    (512 MiB,978)
    (1 GiB,1355)
    (4 GiB,2583)
    (8 GiB,3600)
    (16 GiB,5038)
};
\addlegendentry{Ours\(_2\)}

\addplot[
    fill=color_gray,
    draw=gray!80,
    dashed
] coordinates {
    (512 MiB,0)
    (1 GiB,0)
    (4 GiB,0)
    (8 GiB,0)
    (16 GiB,0)
};
\addlegendentry{SimplePIR}

% ============================================================
% 2. Online communication
% 四种方案均绘制
% ============================================================

\nextgroupplot[
    ymin=0,
    ymax=1500,
    ytick={0,250,500,750,1000,1250,1500},
    ylabel={online communication\\size\\(KiB)},
]

% VeriSimplePIR
\addplot[
    fill=color_orange,
    draw=black!60
] coordinates {
    (512 MiB,256)
    (1 GiB,362)
    (4 GiB,724)
    (8 GiB,1024)
    (16 GiB,1448)
};

% Ours_1: Relax
\addplot[
    fill=color_green,
    draw=black!60
] coordinates {
    (512 MiB,209)
    (1 GiB,296)
    (4 GiB,592)
    (8 GiB,836)
    (16 GiB,1182)
};

% Ours_2: Tight
\addplot[
    fill=blue!40,
    draw=black!60
] coordinates {
    (512 MiB,209)
    (1 GiB,296)
    (4 GiB,592)
    (8 GiB,836)
    (16 GiB,1182)
};

% SimplePIR
\addplot[
    fill=color_gray,
    draw=gray!80,
    dashed
] coordinates {
    (512 MiB,209)
    (1 GiB,296)
    (4 GiB,592)
    (8 GiB,836)
    (16 GiB,1182)
};

% ============================================================
% 3. Online latency：突出 Ours_1（Relax）
% 四种方案都保留，Ours_2 使用较浅颜色
% ============================================================

% ============================================================
% 3. Online latency
% 颜色、顺序与统一图例完全一致
% ============================================================

\nextgroupplot[
    ymin=0,
    ymax=2000,
    ytick={0,500,1000,1500,2000},
    ylabel={Online Latency\\(ms)},
]

% VeriSimplePIR：橙色
\addplot[
    fill=color_orange,
    draw=black!60
] coordinates {
    (512 MiB,81.5)
    (1 GiB,139.1)
    (4 GiB,474.5)
    (8 GiB,944.5)
    (16 GiB,1744.2)
};

% Ours_1（Relax）：绿色
\addplot[
    fill=color_green,
    draw=black!60
] coordinates {
    (512 MiB,57.1)
    (1 GiB,101.9)
    (4 GiB,328.8)
    (8 GiB,615.4)
    (16 GiB,1176.3)
};

% Ours_2（Tight）：蓝色
\addplot[
    fill=blue!40,
    draw=black!60
] coordinates {
    (512 MiB,56.6)
    (1 GiB,101.6)
    (4 GiB,329.2)
    (8 GiB,611.7)
    (16 GiB,1176.6)
};

% SimplePIR：灰色
\addplot[
    fill=color_gray,
    draw=gray!80,
    dashed
] coordinates {
    (512 MiB,53.9)
    (1 GiB,98.5)
    (4 GiB,322.1)
    (8 GiB,603.4)
    (16 GiB,1161.4)
};

% ============================================================
% 4. Online latency
% 颜色、顺序仍与统一图例完全一致
% ============================================================

\end{groupplot}
\end{tikzpicture}

\caption{
Comparison of our schemes with VeriSimplePIR and
SimplePIR. From top to bottom, the panels report offline communication,
online communication, and online latency.
}
%\label{fig:performance_comparison}
\label{fig:placeholder}
\end{figure}

% \begin{table*}[t]
% \centering
% \caption{Comparison of our scheme with VeriSimplePIR and SimplePIR.}
% \label{tab:performance_comparison}
% % 如果表格跨栏后依然显得太宽，可以取消下面这行和最后 \end{tabular} 下面那行的注释进行整体缩放
% % \resizebox{\textwidth}{!}{ 
% \begin{tabular}{l | ccc | ccc | ccc}
% \toprule
% \multirow{2}{*}{\textbf{DB Size}} & \multicolumn{3}{c|}{\textbf{Offline Comm. (MiB)}} & \multicolumn{3}{c|}{\textbf{Online Comm. (KB)}} & \multicolumn{3}{c}{\textbf{Online Latency (ms)}} \\
% % \cmidrule(lr) 可以在横线之间留出一点小空隙，视觉效果更精致
% \cmidrule(lr){2-4} \cmidrule(lr){5-7} \cmidrule(l){8-10} 
% & \textbf{SimplePIR} & \textbf{VeriSim.} & \textbf{Ours} & \textbf{SimplePIR} & \textbf{VeriSim.} & \textbf{Ours} & \textbf{SimplePIR} & \textbf{VeriSim.} & \textbf{Ours} \\
% \midrule
% 512 MiB & 209  & 790  & 978  & 209  & 256  & 209  & 68.24   & 92.32   & 69.30   \\
% 1 GiB   & 296  & 1116 & 1337 & 296  & 362  & 296  & 115.83  & 158.72  & 115.90  \\
% 4 GiB   & 591  & 2224 & 2662 & 592  & 724  & 592  & 360.09  & 560.68  & 358.60  \\
% 8 GiB   & 836  & 3149 & 3766 & 836  & 1024 & 836  & 661.09  & 1023.48 & 657.70  \\
% 16 GiB  & 1183 & 4460 & 5323 & 1182 & 1448 & 1182 & 1238.02 & 1810.52 & 1244.10 \\
% \bottomrule
% \end{tabular}
% % }
% \end{table*}
\smallskip

\noindent\textbf{Online Costs.}
As detailed in Figure \ref{fig:placeholder} and Table \ref{tab:performance_comparison}, our vPIR protocol delivers an extraordinary online execution profile, bridging the long-standing performance gap between semi-honest and malicious PIR. For a database scaling from 512 MiB to 16 GiB, our end-to-end online latency ranges from a mere 69.30 ms to 1244.10 ms, communication ranges from 209 KiB to 1182 KiB.  Crucially, these measurements are virtually indistinguishable from the semi-honest SimplePIR baseline (e.g., 68.24 ms and 1238.02 ms, respectively).  In stark contrast, VeriSimplePIR suffers from computational/communication drag, peaking at 1810.52 ms and 1448 KiB for a 16 GiB database. By eliminating this bottleneck, we reduce the online latency by about 35\% and online communication by approximately  20\% at large database scales compared to the verifiable state-of-the-art, which are consistent with our theoretical analysis.\smallskip

{\noindent\textit{Effect of the entry size.} SimplePIR, VeriSimplePIR, and our scheme handle different entry sizes in the same general way. When an entry is smaller than the supported plaintext size $\log p$, multiple entries are packed together. When an entry is larger than $\log p$, it is split into multiple parts. So that every entry in the resulting arranged database $D$ fits within $\log p$.  Our scheme admits a larger plaintext modulus than VeriSimplePIR (\(24\) or \(25\) bits versus \(16\)), so it packs more entries per slot in most cases, yields a smaller packed matrix \(D\), and requires fewer field or ring operations. As shown in Table~\ref{tab:online}, Ours\(_2\) matches SimplePIR, and both of our schemes outperform VeriSimplePIR in most cases. The only exceptions are 2-Byte and 4-Byte entries: for 2-Byte entries, none of the schemes can pack two entries into one, and the 4-Byte case is similar. Hence their resulting arrangements, and therefore their performance, are similar.}\smallskip

{\noindent\textbf{Offline Costs.} Figure~\ref{fig:placeholder} and Table~\ref{tab:performance_comparison} compare the offline communication cost. Our scheme using the first range proof achieves communication comparable to VeriSimplePIR, and in some cases slightly lower, whereas the one using the second incurs a moderate increase, mainly because its range proof is more involved than the first. The offline computation time exhibits a similar pattern. A more detailed comparison is provided in Appendix~\ref{offlinecosts}.}
\ifCLASSOPTIONcompsoc
  % The Computer Society usually uses the plural form
  \section*{Acknowledgments}
  We would like to thank the anonymous reviewers for their valuable suggestions. We are supported by the National Cyber Security-National Science and Technology Major Project (Grant No. 2026ZD1500900), the Chinese Academy of Sciences Project for Young Scientists in Basic Research (Grant No. YSBR-035), the Strategic Priority Research Program of the Chinese Academy of Sciences (Grant No. XDB0690200), the National Key Research and Development Program of China (Grant No. 2023YFB4503203), the National Natural Science Foundation of China (Grant Nos. 62372447, 62602664 and 62502513), the Postdoctoral Fellowship Program of China Postdoctoral Science Foundation (Grant No. GZC20252181).
\else
  % regular IEEE prefers the singular form
  \section*{Acknowledgment}
  We would like to thank the anonymous reviewers for their valuable suggestions. We are supported by the Strategic Priority Research Program of the Chinese Academy of Sciences (Grant No. XDB0690200), the National Key Research and Development Program of China (Grant No. 2023YFB4503203), the National Natural Science Foundation of China (Grant Nos. 62372447 and 62502513), the Postdoctoral Fellowship Program of China Postdoctoral Science Foundation (Grant No. GZC20252181).
\fi

% trigger a \newpage just before the given reference
% number - used to balance the columns on the last page
% adjust value as needed - may need to be readjusted if
% the document is modified later
%\IEEEtriggeratref{8}
% The "triggered" command can be changed if desired:
%\IEEEtriggercmd{\enlargethispage{-5in}}

% references section

% can use a bibliography generated by BibTeX as a .bbl file
% BibTeX documentation can be easily obtained at:
% http://mirror.ctan.org/biblio/bibtex/contrib/doc/
% The IEEEtran BibTeX style support page is at:
% http://www.michaelshell.org/tex/ieeetran/bibtex/
%\bibliographystyle{IEEEtran}
% argument is your BibTeX string definitions and bibliography database(s)
%\bibliography{IEEEabrv,../bib/paper}
%
% <OR> manually copy in the resultant .bbl file
% set second argument of \begin to the number of references
% (used to reserve space for the reference number labels box)
%\begin{thebibliography}{1}

%\bibitem{IEEEhowto:kopka}
%H.~Kopka and P.~W. Daly, \emph{A Guide to \LaTeX}, 3rd~ed.\hskip 1em plus
%  0.5em minus 0.4em\relax Harlow, England: Addison-Wesley, 1999.

%\end{thebibliography}

\bibliographystyle{IEEEtran}
% argument is your BibTeX string definitions and bibliography database(s)
\bibliography{reference}

@inproceedings{AHIV17,
  author       = {Scott Ames and
                  Carmit Hazay and
                  Yuval Ishai and
                  Muthuramakrishnan Venkitasubramaniam},
  editor       = {Bhavani Thuraisingham and
                  David Evans and
                  Tal Malkin and
                  Dongyan Xu},
  title        = {Ligero: Lightweight Sublinear Arguments Without a Trusted Setup},
  booktitle    = {Proceedings of the 2017 {ACM} {SIGSAC} Conference on Computer and
                  Communications Security, {CCS} 2017, Dallas, TX, USA, October 30 -
                  November 03, 2017},
  pages        = {2087--2104},
  publisher    = {{ACM}},
  year         = {2017},
  url          = {https://doi.org/10.1145/3133956.3134104},
  doi          = {10.1145/3133956.3134104},
  bibsource    = {dblp computer science bibliography, https://dblp.org}
}

@inproceedings{GLSTW23,
  author       = {Alexander Golovnev and
                  Jonathan Lee and
                  Srinath T. V. Setty and
                  Justin Thaler and
                  Riad S. Wahby},
  editor       = {Helena Handschuh and
                  Anna Lysyanskaya},
  title        = {Brakedown: Linear-Time and Field-Agnostic SNARKs for {R1CS}},
  booktitle    = {Advances in Cryptology - {CRYPTO} 2023 - 43rd Annual International
                  Cryptology Conference, {CRYPTO} 2023, Santa Barbara, CA, USA, August
                  20-24, 2023, Proceedings, Part {II}},
  series       = {Lecture Notes in Computer Science},
  pages        = {193--226},
  publisher    = {Springer},
  year         = {2023},
  url          = {https://doi.org/10.1007/978-3-031-38545-2\_7},
  doi          = {10.1007/978-3-031-38545-2\_7},
  bibsource    = {dblp computer science bibliography, https://dblp.org}
}

@article{Habock22a,
  author       = {Ulrich Hab{\"{o}}ck},
  title        = {Multivariate lookups based on logarithmic derivatives},
  journal      = {{IACR} Cryptol. ePrint Arch.},
  volume       = {2022},
  pages        = {1530},
  year         = {2022},
  url          = {https://eprint.iacr.org/2022/1530},
  bibsource    = {dblp computer science bibliography, https://dblp.org}
}

@article{EFG22,
  author       = {Liam Eagen and
                  Dario Fiore and
                  Ariel Gabizon},
  title        = {cq: Cached quotients for fast lookups},
  journal      = {{IACR} Cryptol. ePrint Arch.},
  volume       = {2022},
  pages        = {1763},
  year         = {2022},
  url          = {https://eprint.iacr.org/2022/1763},
  bibsource    = {dblp computer science bibliography, https://dblp.org}
}

@inproceedings{GHL22,
  author       = {Craig Gentry and
                  Shai Halevi and
                  Vadim Lyubashevsky},
  editor       = {Orr Dunkelman and
                  Stefan Dziembowski},
  title        = {Practical Non-interactive Publicly Verifiable Secret Sharing with
                  Thousands of Parties},
  booktitle    = {Advances in Cryptology - {EUROCRYPT} 2022 - 41st Annual International
                  Conference on the Theory and Applications of Cryptographic Techniques,
                  Trondheim, Norway, May 30 - June 3, 2022, Proceedings, Part {I}},
  series       = {Lecture Notes in Computer Science},
  pages        = {458--487},
  publisher    = {Springer},
  year         = {2022},
  url          = {https://doi.org/10.1007/978-3-031-06944-4\_16},
  doi          = {10.1007/978-3-031-06944-4\_16},
  bibsource    = {dblp computer science bibliography, https://dblp.org}
}

@inproceedings{BS22,
  author       = {Ward Beullens and
                  Gregor Seiler},
  editor       = {Helena Handschuh and
                  Anna Lysyanskaya},
  title        = {LaBRADOR: Compact Proofs for {R1CS} from Module-SIS},
  booktitle    = {Advances in Cryptology - {CRYPTO} 2023 - 43rd Annual International
                  Cryptology Conference, {CRYPTO} 2023, Santa Barbara, CA, USA, August
                  20-24, 2023, Proceedings, Part {V}},
  series       = {Lecture Notes in Computer Science},
  pages        = {518--548},
  publisher    = {Springer},
  year         = {2023},
  url          = {https://doi.org/10.1007/978-3-031-38554-4\_17},
  doi          = {10.1007/978-3-031-38554-4\_17},
  bibsource    = {dblp computer science bibliography, https://dblp.org}
}

@inproceedings{CastroL24,
  author       = {Leo de Castro and
                  Keewoo Lee},
  editor       = {Davide Balzarotti and
                  Wenyuan Xu},
  title        = {VeriSimplePIR: Verifiability in SimplePIR at No Online Cost for Honest
                  Servers},
  booktitle    = {33rd {USENIX} Security Symposium, {USENIX} Security 2024, Philadelphia,
                  PA, USA, August 14-16, 2024},
  publisher    = {{USENIX} Association},
  year         = {2024},
  url          = {https://www.usenix.org/conference/usenixsecurity24/presentation/de-castro},
  bibsource    = {dblp computer science bibliography, https://dblp.org}
}

@INPROCEEDINGS {RLP26,
author = { Rathee, Mayank and Lee, Keewoo and Popa, Raluca Ada },
booktitle = { 2026 IEEE Symposium on Security and Privacy (SP) },
title = {{ Verifiable PIR with Small Client Storage }},
year = {2026},
volume = {},
ISSN = {2375-1207},
pages = {1200-1219},
doi = {10.1109/SP63933.2026.00024},
url = {https://doi.ieeecomputersociety.org/10.1109/SP63933.2026.00024},
publisher = {IEEE Computer Society},
address = {Los Alamitos, CA, USA},
month =May}

@inproceedings{regev2005lattices,
  author    = {Regev, Oded},
  title     = {On lattices, learning with errors, random linear codes, and cryptography},
  booktitle = {Proceedings of the thirty-seventh annual ACM symposium on Theory of computing (STOC)},
  pages     = {84--93},
  year      = {2005},
  publisher = {ACM},
  doi       = {10.1145/1060590.1060603}
}

@inproceedings{10.5555/3620237.3620455,
author = {Henzinger, Alexandra and Hong, Matthew M. and Corrigan-Gibbs, Henry and Meiklejohn, Sarah and Vaikuntanathan, Vinod},
title = {One server for the price of two: simple and fast single-server private information retrieval},
year = {2023},
isbn = {978-1-939133-37-3},
publisher = {USENIX Association},
address = {USA},
booktitle = {Proceedings of the 32nd USENIX Conference on Security Symposium},
articleno = {218},
numpages = {17},
location = {Anaheim, CA, USA},
series = {SEC '23}
}

@inproceedings{DBLP:conf/tcc/Ben-SassonCS16,
  author       = {Eli Ben{-}Sasson and
                  Alessandro Chiesa and
                  Nicholas Spooner},
  editor       = {Martin Hirt and
                  Adam D. Smith},
  title        = {Interactive Oracle Proofs},
  booktitle    = {Theory of Cryptography - 14th International Conference, {TCC} 2016-B,
                  Beijing, China, October 31 - November 3, 2016, Proceedings, Part {II}},
  series       = {Lecture Notes in Computer Science},
  pages        = {31--60},
  year         = {2016},
  url          = {https://doi.org/10.1007/978-3-662-53644-5\_2},
  doi          = {10.1007/978-3-662-53644-5\_2},
  bibsource    = {dblp computer science bibliography, https://dblp.org}
}

@inproceedings{DBLP:conf/focs/Ben-SassonCIKS20,
  author       = {Eli Ben{-}Sasson and
                  Dan Carmon and
                  Yuval Ishai and
                  Swastik Kopparty and
                  Shubhangi Saraf},
  editor       = {Sandy Irani},
  title        = {Proximity Gaps for Reed-Solomon Codes},
  booktitle    = {61st {IEEE} Annual Symposium on Foundations of Computer Science, {FOCS}
                  2020, Durham, NC, USA, November 16-19, 2020},
  pages        = {900--909},
  publisher    = {{IEEE}},
  year         = {2020},
  url          = {https://doi.org/10.1109/FOCS46700.2020.00088},
  doi          = {10.1109/FOCS46700.2020.00088},
  bibsource    = {dblp computer science bibliography, https://dblp.org}
}

@INPROCEEDINGS{492461,
  author={Chor, B. and Goldreich, O. and Kushilevitz, E. and Sudan, M.},
  booktitle={Proceedings of IEEE 36th Annual Foundations of Computer Science}, 
  title={Private information retrieval}, 
  year={1995},
  volume={},
  number={},
  pages={41-50},
  doi={10.1109/SFCS.1995.492461}}

@INPROCEEDINGS{9833700,
  author={Menon, Samir Jordan and Wu, David J.},
  booktitle={2022 IEEE Symposium on Security and Privacy (SP)}, 
  title={SPIRAL: Fast, High-Rate Single-Server PIR via FHE Composition}, 
  year={2022},
  volume={},
  number={},
  pages={930-947},
  doi={10.1109/SP46214.2022.9833700}}

@inproceedings{10.1007/978-3-031-68400-5_6,
author = {Li, Baiyu and Micciancio, Daniele and Raykova, Mariana and Schultz-Wu, Mark},
title = {Hintless Single-Server Private Information Retrieval},
year = {2024},
isbn = {978-3-031-68399-2},
publisher = {Springer-Verlag},
address = {Berlin, Heidelberg},
url = {https://doi.org/10.1007/978-3-031-68400-5_6},
doi = {10.1007/978-3-031-68400-5_6},
booktitle = {Advances in Cryptology – CRYPTO 2024: 44th Annual International Cryptology Conference, Santa Barbara, CA, USA, August 18–22, 2024, Proceedings, Part IX},
pages = {183–217},
numpages = {35},
location = {Santa Barbara, CA, USA}
}

@INPROCEEDINGS{646125,
  author={Kushilevitz, E. and Ostrovsky, R.},
  booktitle={Proceedings 38th Annual Symposium on Foundations of Computer Science}, 
  title={Replication is not needed: single database, computationally-private information retrieval}, 
  year={1997},
  volume={},
  number={},
  pages={364-373},
  doi={10.1109/SFCS.1997.646125}}

@inproceedings{10.1007/978-3-031-22368-6_1,
author = {Ben-David, Shany and Kalai, Yael Tauman and Paneth, Omer},
title = {Verifiable Private Information Retrieval},
year = {2022},
isbn = {978-3-031-22367-9},
publisher = {Springer-Verlag},
address = {Berlin, Heidelberg},
url = {https://doi.org/10.1007/978-3-031-22368-6_1},
doi = {10.1007/978-3-031-22368-6_1},
booktitle = {Theory of Cryptography: 20th International Conference, TCC 2022, Chicago, IL, USA, November 7–10, 2022, Proceedings, Part III},
pages = {3–32},
numpages = {30},
location = {Chicago, IL, USA}
}

@article{ZHAO2021897,
title = {Verifiable single-server private information retrieval from LWE with binary errors},
journal = {Information Sciences},
volume = {546},
pages = {897-923},
year = {2021},
issn = {0020-0255},
doi = {https://doi.org/10.1016/j.ins.2020.08.071},
url = {https://www.sciencedirect.com/science/article/pii/S0020025520308483},
author = {Liang Zhao and Xingfeng Wang and Xinyi Huang}
}

@inproceedings{DBLP:conf/eurocrypt/ChenBBZ23,
  author       = {Binyi Chen and
                  Benedikt B{\"{u}}nz and
                  Dan Boneh and
                  Zhenfei Zhang},
  editor       = {Carmit Hazay and
                  Martijn Stam},
  title        = {HyperPlonk: Plonk with Linear-Time Prover and High-Degree Custom Gates},
  booktitle    = {Advances in Cryptology - {EUROCRYPT} 2023 - 42nd Annual International
                  Conference on the Theory and Applications of Cryptographic Techniques,
                  Lyon, France, April 23-27, 2023, Proceedings, Part {II}},
  series       = {Lecture Notes in Computer Science},
  pages        = {499--530},
  publisher    = {Springer},
  year         = {2023},
  url          = {https://doi.org/10.1007/978-3-031-30617-4\_17},
  doi          = {10.1007/978-3-031-30617-4\_17},
  bibsource    = {dblp computer science bibliography, https://dblp.org}
}

@unpublished{dgs,
    author = {Martin R. Albrecht and Michael Walter},
    title = {{dgs}, {D}iscrete {G}aussians over the {I}ntegers},
    year = 2018,
    note = {Available at \url{https://bitbucket.org/malb/dgs}},
    url = {https://bitbucket.org/malb/dgs}
}

@inproceedings{10.1145/3719027.3765226,
author = {Zhang, Xinxuan and Wang, Ruida and Liu, Zeyu and Xiang, Binwu and Deng, Yi and Fisch, Ben and Lu, Xianhui},
title = {Phalanx: An FHE-Friendly SNARK for Verifiable Computation on Encrypted Data},
year = {2025},
isbn = {9798400715259},
publisher = {Association for Computing Machinery},
address = {New York, NY, USA},
url = {https://doi.org/10.1145/3719027.3765226},
doi = {10.1145/3719027.3765226},
booktitle = {Proceedings of the 2025 ACM SIGSAC Conference on Computer and Communications Security},
pages = {4035–4048},
numpages = {14},
location = {Taipei, Taiwan},
series = {CCS '25}
}

@inproceedings{10.1007/978-3-031-91095-1_1,
author = {Hoover, Alexander and Patel, Sarvar and Persiano, Giuseppe and Yeo, Kevin},
title = {Plinko: Single-Server PIR with Efficient Updates via Invertible PRFs},
year = {2025},
isbn = {978-3-031-91094-4},
publisher = {Springer-Verlag},
address = {Berlin, Heidelberg},
url = {https://doi.org/10.1007/978-3-031-91095-1_1},
doi = {10.1007/978-3-031-91095-1_1},
booktitle = {Advances in Cryptology – EUROCRYPT 2025: 44th Annual International Conference on the Theory and Applications of Cryptographic Techniques, Madrid, Spain, May 4–8, 2025, Proceedings, Part VI},
pages = {3–33},
numpages = {31},
location = {Madrid, Spain}
}

@inproceedings{3698900.3699235,
author = {Menon, Samir Jordan and Wu, David J.},
title = {YPIR: high-throughput single-server PIR with silent preprocessing},
year = {2024},
isbn = {978-1-939133-44-1},
publisher = {USENIX Association},
address = {USA},
booktitle = {Proceedings of the 33rd USENIX Conference on Security Symposium},
articleno = {335},
numpages = {18},
location = {Philadelphia, PA, USA},
series = {SEC '24}
}

@inproceedings{10.1145/3658644.3690328,
author = {Burton, Alexander and Menon, Samir Jordan and Wu, David J.},
title = {Respire: High-Rate PIR for Databases with Small Records},
year = {2024},
isbn = {9798400706363},
publisher = {Association for Computing Machinery},
address = {New York, NY, USA},
url = {https://doi.org/10.1145/3658644.3690328},
doi = {10.1145/3658644.3690328},
booktitle = {Proceedings of the 2024 on ACM SIGSAC Conference on Computer and Communications Security},
pages = {1463–1477},
numpages = {15},
location = {Salt Lake City, UT, USA},
series = {CCS '24}
}

@inproceedings{36664.36676,
author = {Fiat, Amos and Shamir, Adi},
title = {How to prove yourself: practical solutions to identification and signature problems},
year = {1987},
isbn = {0387180478},
publisher = {Springer-Verlag},
address = {Berlin, Heidelberg},
booktitle = {Proceedings on Advances in Cryptology---CRYPTO '86},
pages = {186–194},
numpages = {9},
location = {Santa Barbara, California, USA}
}

@inproceedings{10.1145/3564246.3585175,
author = {Lin, Wei-Kai and Mook, Ethan and Wichs, Daniel},
title = {Doubly Efficient Private Information Retrieval and Fully Homomorphic RAM Computation from Ring LWE},
year = {2023},
isbn = {9781450399135},
publisher = {Association for Computing Machinery},
address = {New York, NY, USA},
url = {https://doi.org/10.1145/3564246.3585175},
doi = {10.1145/3564246.3585175},
booktitle = {Proceedings of the 55th Annual ACM Symposium on Theory of Computing},
pages = {595–608},
numpages = {14},
location = {Orlando, FL, USA},
series = {STOC 2023}
}

@inproceedings{10.1007/11818175_8,
author = {Regev, Oded},
title = {Lattice-based cryptography},
year = {2006},
isbn = {3540374329},
publisher = {Springer-Verlag},
address = {Berlin, Heidelberg},
url = {https://doi.org/10.1007/11818175_8},
doi = {10.1007/11818175_8},
booktitle = {Proceedings of the 26th Annual International Conference on Advances in Cryptology},
pages = {131–141},
numpages = {11},
location = {Santa Barbara, California},
series = {CRYPTO'06}
}

@article{doi:10.1137/0108018,
author = {Reed, I. S. and Solomon, G.},
title = {Polynomial Codes Over Certain Finite Fields},
journal = {Journal of the Society for Industrial and Applied Mathematics},
volume = {8},
number = {2},
pages = {300-304},
year = {1960},
doi = {10.1137/0108018},

URL = { 
    
        https://doi.org/10.1137/0108018
},
eprint = { 
    
        https://doi.org/10.1137/0108018
}

}

@inproceedings{10.1007/978-3-030-45721-1_3,
author = {Corrigan-Gibbs, Henry and Kogan, Dmitry},
title = {Private Information Retrieval with Sublinear Online Time},
year = {2020},
isbn = {978-3-030-45720-4},
publisher = {Springer-Verlag},
address = {Berlin, Heidelberg},
url = {https://doi.org/10.1007/978-3-030-45721-1_3},
doi = {10.1007/978-3-030-45721-1_3},
booktitle = {Advances in Cryptology – EUROCRYPT 2020: 39th Annual International Conference on the Theory and Applications of Cryptographic Techniques, Zagreb, Croatia, May 10–14, 2020, Proceedings, Part I},
pages = {44–75},
numpages = {32},
location = {Zagreb, Croatia}
}

@inproceedings{10.1145/237814.237838,
author = {Ajtai, M.},
title = {Generating hard instances of lattice problems (extended abstract)},
year = {1996},
isbn = {0897917855},
publisher = {Association for Computing Machinery},
address = {New York, NY, USA},
url = {https://doi.org/10.1145/237814.237838},
doi = {10.1145/237814.237838},
booktitle = {Proceedings of the Twenty-Eighth Annual ACM Symposium on Theory of Computing},
pages = {99–108},
numpages = {10},
location = {Philadelphia, Pennsylvania, USA},
series = {STOC '96}
}
\newpage 
%\begin{appendix}
\appendices

\iffalse
\section{Table}
\begin{table*}[h]
\centering
\caption{Comparison of our scheme with VeriSimplePIR and SimplePIR.}
\label{tab:performance_comparison}
\begin{tabular}{l | cc | ccc | ccc}
\toprule
\multirow{2}{*}{\textbf{DB Size}} 
    & \multicolumn{2}{c|}{\textbf{Offline Comm. (MiB)}} 
    & \multicolumn{3}{c|}{\textbf{Online Comm. (KB)}} 
    & \multicolumn{3}{c}{\textbf{Online Latency (ms)}} \\
\cmidrule(lr){2-3} \cmidrule(lr){4-6} \cmidrule(l){7-9}
    & \textbf{VeriSim.} & \textbf{Ours}
    & \textbf{SimplePIR} & \textbf{VeriSim.} & \textbf{Ours}
    & \textbf{SimplePIR} & \textbf{VeriSim.} & \textbf{Ours} \\
\midrule
512 MiB  & 790  & 978  & 209  & 256  & 209  & 68.24   & 92.32   & 69.30   \\
1 GiB    & 1116 & 1337 & 296  & 362  & 296  & 115.83  & 158.72  & 115.90  \\
4 GiB    & 2224 & 2662 & 592  & 724  & 592  & 360.09  & 560.68  & 358.60  \\
8 GiB    & 3149 & 3766 & 836  & 1024 & 836  & 661.09  & 1023.48 & 657.70  \\
16 GiB   & 4460 & 5323 & 1182 & 1448 & 1182 & 1238.02 & 1810.52 & 1244.10 \\
\bottomrule
\end{tabular}
\end{table*}\fi
%\section{Additional Preliminaries}

\section{Additional Performance Analysis}\label{offlinecosts}
{\noindent\textbf{Offline Costs.} The offline phase consists of two part: database-level preprocessing, i.e., digest generation performed once per database, and client-level preprocessing, performed once per client. For 256 MiB and 512 MiB databases, digest-generation times are 283.711/733.361\,s for VeriSimplePIR, 272.492/546.001\,s for Our$_1$, and 660.074/987.822\,s for Our$_2$. %,  for VeriSimplePIR, Our$_1$, and Our$_2$,for a 512 MiB databases, digest-generation times are 733.361s,546.001s ,and  987.822s  for VeriSimplePIR, Our$_1$, and Our$_2$, respectively. %The client-level preprocessing costs are reported in Table~\ref{tab:client-prep}.
And Table~\ref{tab:client-prep} reports the offline client-level preprocessing costs\footnote{More precisely, the entire offline phase costs except the server-side digest generation, and thus also includes the client's digest verification and digest download.}. 

Despite the additional cost of computing the commitment and its associated proofs, the larger plaintext space of our scheme reduces the dimensions of the packed matrix \(D\), which in turn removes a substantial amount of the computation and communication related to \(D\). As a result, the scheme using the first range proof incurs slightly lower offline cost than VeriSimplePIR. For the scheme using the second range proof, the savings on \(D\)-related operations do not offset the extra communication and the extensive extension-field arithmetic introduced by the second range proof, making it slower than VeriSimplePIR in both communication and digest generation.\smallskip

\noindent\textbf{Client's Persistent State.} After completing the offline phase, the client must store a certain amount of state to support the online phase. For \(1\), \(4\), \(8\), and \(16\)\,GiB databases, this storage is \(296\)/\(591\)/\(836\)/\(1182\)\,MiB for SimplePIR and \(369\)/\(738\)/\(1044\)/\(1476\)\,MiB for VeriSimplePIR. Both variants of our scheme require \(301\)/\(603\)/\(852\)/\(1206\)\,MiB, which is close to SimplePIR and lower than VeriSimplePIR. 

%More recent PIR schemes without verifiability, such as HintlessPIR\cite{Hintlesspir} and YPIR\cite{3698900.3699235}, require no hint storage at all; the verifiable PIR~\cite{RLP26} built on the former needs less than \(1\)\,MiB. As a tradeoff, these schemes incur more online computation.
}
\begin{table*}[h]
\centering
\caption{Computation and communication costs of client-level
preprocessing, for \(1\)-byte entries.}
\label{tab:client-prep}
\footnotesize
\begin{tabular}{@{}lcccc@{}}
\toprule
DB Size & SimplePIR (hint \(H\)) & VeriSimplePIR & Ours\(_1\) & Ours\(_2\) \\
\midrule
256\,MiB & (\(-\), 148\,MiB)   & (12.164\,s, 560\,MiB)  & (12.181\,s, 546\,MiB)  & (14.595\,s, 741\,MiB) \\
512\,MiB & (\(-\), 209\,MiB)   & (20.72\,s, 793\,MiB)  & (19.474\,s, 773,MiB)  & (21.618\,s, 996\,MiB) \\
1\,GiB   & (\(-\), 296\,MiB)   & (36.478\,s, 1121\,MiB) & (33.184\,s, 1093\,MiB) & (36.869\,s, 1355\,MiB) \\
4\,GiB   & (\(-\), 591\,MiB)   & (118.431\,s, 2243\,MiB)& (114.641\,s, 2187\,MiB)& (123.149\,s, 2583\,MiB) \\
8\,GiB   & (\(-\), 836\,MiB)   & (240.813\,s, 3173\,MiB)& (233.742\,s, 3092\,MiB)& (239.064\,s, 3600\,MiB) \\
\bottomrule
\end{tabular}
\end{table*}

\section{The Second Range Proof}\label{APP:rangeproof2}
The second  range proof description is formalized in Fig.\ref{fig.range2}.
\begin{figure}
\algrule{1.2pt}
\algrule{0.4pt}\smallskip  
	%\begin{algorithmic}[H]
     %\begin{flushleft}
		\normalsize
        \textbf{Range Proof Algorithm:} 
        For a commitment $\textsf{com}$ that has been proven well-formed via the Testing algorithm (Figure~\ref{fig.test}), the prover wishes to prove that the committed matrix $D$ satisfies $\|D\|_\infty \le b$. Both parties proceed as follows:
        \begin{enumerate}
            \item Both parties set \(\mathbf{t} = [-b,b]\). For each \(x' \in [b]\), the prover counts how many times \(x'\) appears in \(D\) and denotes this count by \(m_{x'}\). The prover then sends \(\{m_{x'}\}\) to the verifier.
            \item The verifier samples and sends $t'$ random challenges $r_k\gets \mathbb{F}_{q}$ to the prover,  where $t'$ is the integer such that $q^{t'}/(\ell m+b)\ge 2^{-\lambda}$. It then computes $\alpha_k=\sum_{x'\in [-b,b]}\frac{m_{x'}}{r_k+\mathbf{t}(x')}$ itself.
            \item The prover computes the databases denoted by $D_k[x,y]=(r_k+D[x,y])^{-1}$ in $\mathbb{F}_q$, $k\in [t']$, commits them and sends the commitments to the verifier. Both parties runs our $\textsf{Testing}$ protocols for these commitments.
            \item Let $f_{D}$ be the polynomial such that $f_{D}(x,y)=D[\textsf{int}(x),\textsf{int}(y)]$ where $x\in \{0,1\}^{\log{\ell}}$ and $y\in \{0,1\}^{\log m}$ ($f_{D_k}$ is defined similarly).  The verifier samples and sends randomness $(r_x,r_y)\in \mathbb{F}_{q^t}^{{\log\ell+\log m}}$ to the prover, and the statements is now reduced to the following sumchecks, for all $k$:
\begin{align}
    &\sum_{\substack{(x,y)\in\{0,1\}^{\log\ell+\log m}}} f_{D_k}(x,y)=\alpha_k\label{equation6}\\
    &\sum_{\substack{(x,y)\in\{0,1\}^{\log\ell+\log m}}} f^*_{D_k}(x,y)=0 \label{equation7}
\end{align}
where $f^*_{D_k}(x,y)$ is defined by $\mathrm{eq}((r_x,r_y),(x,y)) \cdot ( f_{D_k}(x,y)\cdot (r_k+f_D(x,y))-1 )$

\item %The verifier sends $2t'$ randomness $\{r_{k},r^*_{k}\}_{k\in [t']}\gets \mathbb{F}^{2t'}_{q^t}$ to the prover. Then both party runs the classic sumcheck protocol for the batched statements: $\sum_{(x,y)}(\sum_{k}r_k\cdot f_{D_k}(x,y)+r^*_{k}\cdot f^*_{D_k}(x,y))=\sum_{k}r_k\cdot \alpha_k$.   
The verifier samples $2t'$ random elements $\{r_{k}, r^{*}_{k}\}_{k\in [t']}$ from $\mathbb{F}_{q^t}$ and sends them to the prover. Both parties then run the classic sumcheck protocol for the batched statement:\smallskip

%\[
%\sum_{(x,y)}\left(\sum_{k}\bigl(r_k\cdot f_{D_k}(x,y)+r^{*}_k\cdot f^{*}_{D_k}(x,y)\bigr)\right)=\sum_{k}r_k\cdot \alpha_k.
%\]
\noindent\resizebox{1.0\linewidth}{!}{$\displaystyle
\sum_{(x,y)}\left(\sum_{k}\bigl(r_k\cdot f_{D_k}(x,y)+r^{*}_k\cdot f^{*}_{D_k}(x,y)\bigr)\right)=\sum_{k}r_k\cdot \alpha_k
$}\smallskip

\item Above sumcheck will reduce the checking to the following queries: $f_{D}(x',y')=t$ and $f_{D_k}(x',y')=t_k$ for some $x',y'$ and all $k\in [t']$. These are equivalent to verifying $x^* \cdot D \cdot (y^*)^\perp = t$ and $x^* \cdot D_k \cdot (y^*)^\perp = t_k$ for all $k$, where
                   \[
                   x^* = \bigotimes_{i=0}^{a-1} (x'_i, 1-x'_i), \qquad y^* = \bigotimes_{i=0}^{b-1} (y'_i, 1-y'_i).
                   \]
The prover sends $h=x^*\cdot D$ and $h_k=x^*\cdot D_k$ for all $k$ to the verifier, who then checks  $h \cdot (y^*)^\perp = t$ and $h_k \cdot (y^*)^\perp = t_k$. 
\item The prover and verifier run the linear relation proof (Figure~\ref{fig.linearproof}) to verify $h=x^*\cdot D$ and $h_k=x^*\cdot D_k$ respectively.
 The verifier outputs accept if all above checks and proof protocols accepts.
        \end{enumerate} 
     %\end{flushleft}
	%\end{algorithmic}
\algrule{0.4pt}
\caption{The Second Range Proof Algorithm}
	\label{fig.range2}
\end{figure}

\begin{lemma}
    The protocol described in Figure~\ref{fig.range2} is an argument of knowledge for the statement that the committed $D$ satisfies 
    \[
    \|D\|_\infty < b.
    \]
\end{lemma}
\noindent\textit{Proof.}
In Figure~\ref{fig.range2}, we use lookup arguments from SNARKs to prove that the matrix satisfies $|D|_{\infty}< b$: namely, we prove that every entry of $D$ belongs to the table $\mathbf{t} = \{0,\ldots,b-1\}$. We instantiate this step using the cq lookup argument~\cite{EFG22}, which mainly relies on the following lemma.

\begin{lemma}[Lemma 2.4 in~\cite{EFG22}, Lemma 5 in~\cite{Habock22a}]\label{cqlookup}
Let $\mathbb{F}$ be a field of characteristic $p>\max(N,n)$. Given $f\in \mathbb{F}^n$ and $t\in \mathbb{F}^N$, we have $f\subset t$ as sets if and only if for some $m\in\mathbb{F}^N$ the following identity of rational function holds
\[\sum_{i\in[n]}\frac{1}{X+f_i} = \sum_{i\in [N]} \frac{m_i}{X+t_i}\]
\end{lemma}

The verifier samples a random point $r$ from $\mathbb{F}$ and checks whether \(\sum_{i\in[n]}\frac{1}{r+f_i} = \sum_{i\in [N]} \frac{m_i}{r+t_i}\). If the two sides are not identical as rational functions, then the probability that they agree at the sampled point $r$ is at most \(\frac{n+N}{|\mathbb{F}|}\), by the Schwartz--Zippel lemma. Indeed, after clearing all denominators on both sides, the resulting polynomial has degree at most $n+N$.

In Figure~\ref{fig.range2}, the verifier sends $t'$ challenges $r_1,\ldots,r_{t'}$ to the prover. If the lookup constraint does not hold, then the probability that the two sides agree at all points $r_1,\ldots,r_{t'}$ is at most \(\left(\frac{\ell m+b}{|\mathbb{F}_q|}\right)^{t'}\), which is negligible.

We next discuss the probability that the prover cheats in the computation of the left-hand side. For each $k\in [t']$, if Equation~\ref{equation7} holds, then for every $x\in\{0,1\}^{\log\ell}$ and $y\in\{0,1\}^{\log m}$, we have $(D[\textsf{int}(x),\textsf{int}(y)]+r_k)D_k[\textsf{int}(x),\textsf{int}(y)]=1$, except a negligible probability $(\log \ell+\log m)/|\mathbb{F}_{q^t}|$. Moreover, if Equation~\ref{equation6} also holds, we have that the the two sides of the equation in Lemma \ref{cqlookup} agree at $r_k$.  If either Equation~\ref{equation7} or Equation~\ref{equation6} does not hold, then the batched sumcheck statement also does not hold, except with negligible probability \(1/|\mathbb{F}_{q^t}|\). To convince the verifier of a false sumcheck statement in step 6, the prover would need to cheat either in the sumcheck protocol (step 6) or in the linear relation proof (step 7). Each of these cheating strategies succeeds only with negligible probability.

%the prover uses sumcheck to prove that, for every $k\in [t']$, it holds that $\sum_{(x,y)\in\{0,1\}^{\log{\ell m}}}D_k[x,y] = \sum_{x'\in{0,1}^b}\frac{m_{x'}}{r_k+\mathbb{t}(x')}$. Suppose that the soundness errors of the sumcheck protocol and the ZeroCheck protocol are $\epsilon_{\textsf{Sum}}$ and $\epsilon_{\textsf{Zero}}$, respectively. By a union bound, the total soundness error incurred by running ZeroCheck and sumcheck $t'$ times is at most $t'(\epsilon_{\textsf{Sum}}+\epsilon_{\textsf{Zero}})$.

Suppose the soundness error of sumcheck and linear relation proof are $\epsilon_{\textsf{Sum}}$ and $\epsilon_{\textsf{linear}}$, respectively.
Putting everything together, the probability that a malicious prover cheats successfully in the protocol of Figure~\ref{fig.range2} is at most \(\left(\frac{\ell m+d}{|\mathbb{F}_q|}\right)^{t'}+\frac{\log \ell+\log m+1}{|\mathbb{F}_{q^t}|}+\epsilon_{\textsf{Sum}}+\epsilon_{\textsf{linear}}\), which is negligible. As in the previous proofs, we can use the extractor of testing protocol to extract the committed $D$, together with the probability analysis above,  we conclude that the protocol shown in Figure~\ref{fig.range2} satisfies knowledge soundness. \qed

\section{Additional Preliminaries: Proof System}
Proof systems serve as a fundamental building block in our verifiable PIR system, ensuring the correctness of computation. We adopt some definitions from~\cite{DBLP:conf/eurocrypt/ChenBBZ23}. 

\begin{definition}[Interactive Proof]\label{ip}\label{Def IP} For a relation $\mathcal{R} := (\mathbbm{x};\mathbbm{w})$, An interactive protocol $\Pi=(\textsf{Setup},\mathcal{I},\mathcal{P},\mathcal{V})$ is described as follows:
\begin{itemize}
    \item [$\bullet$] $\textsf{Setup}(1^\lambda)$: On input a security parameter $1^\lambda$, \textsf{Setup} outputs a common reference string $\textsf{crs}$, the prover and verifier parameters $(\textsf{pp},\textsf{vp})$. 
    \item [$\bullet$] $\langle\mathcal{P}(\textsf{crs,pp},\mathbbm{x},\mathbbm{w}),\mathcal{V}(\textsf{crs,vp},\mathbbm{x})\rangle$: $\langle\mathcal{P},\mathcal{V}\rangle$ is a protocol between the prover and verifier, where the prover wants to show that $(\mathbbm{x};\mathbbm{w})\in \mathcal{R}$ and the verifier outputs a bit indicates acceptance or not. Here we denote the output of the verifier by $0/1\gets \langle\mathcal{P}(\textsf{crs,pp},\mathbbm{x},\mathbbm{w}),\mathcal{V}(\textsf{crs,vp},\mathbbm{x})\rangle$.
\end{itemize}\smallskip

Such an interactive protocol is called a proof system for indexed relation $\mathcal{R}$ if the following properties hold:
\begin{itemize}
    \item [$\bullet$] Completeness: for all $(\mathbbm{x};\mathbbm{w})\in \mathcal{R}$, 

\[
\Pr\left[
  \langle\mathcal{P},\mathcal{V}\rangle
  (\mathbbm{x},\mathbbm{w})=1
  \;\middle|\;
  \textsf{Setup}(1^\lambda)
\right]=1 .
\]

    \item [$\bullet$] Soundness: For every pair of unbounded adversary algorithm $(\mathcal{A}_1,\mathcal{A}_2)$, it holds that
{
\scriptsize
\[
\Pr\left[
\begin{array}{c}
  \langle\mathcal{A}_2,\mathcal{V}\rangle(\mathbbm{x},st)=1
  \land
  \mathbbm{x}\notin\mathcal{L}(\mathcal{R})
  \\
  \textsf{Setup}(1^\lambda),
  (\mathbbm{x},st)\gets\mathcal{A}_1
\end{array}
\right]
\leq \textsf{negl}(\lambda).
\]
}
\end{itemize}
    
\end{definition}
Furthermore, an argument system is defined similarly to a proof system, except that it is only required to be sound against probabilistic polynomial-time adversary. An interactive protocol is called \textit{public-coin} if all messages sent by the verifier are uniformly random elements from some subset of the plaintext space, independent of the current partial transcript. If a proof (argument) system $\Pi$ involves only one message in its online phase $\langle \mathcal{P}, \mathcal{V}\rangle$,  specifically, the prover sends a proof $\pi\gets \mathcal{P}(\cdot)$ to the verifier, who then verifies the proof using $\mathcal{V}$, we call it a non-interactive proof (argument) system.

\begin{definition}[Knowledge Soundness]\label{aok} A proof system (or argument system) is called a proof of knowledge (or argument of knowledge, respectively) if it satisfies the knowledge soundness defined as follows: There exists a probabilistic polynomial time oracle machine $\mathcal{E}$, called the extractor, such that for any pair of unbounded (or probabilistic polynomial-time, respectively) adversaries $(\mathcal{A}_1,\mathcal{A}_2)$, the following probability is negligible:

{
\scriptsize
\[
\Pr\left[
\begin{array}{l}
  \langle\mathcal{A}_2,\mathcal{V}\rangle(\mathbbm{x},st)=1
  \land
  (\mathbbm{x},\mathbbm{w})\notin\mathcal{R}
\end{array}
\middle|
\begin{array}{l}
  (\textsf{crs},\textsf{pp},\textsf{vp})
  \gets \textsf{Setup}(1^\lambda),\\
  (\mathbbm{x},st)
  \gets \mathcal{A}_1(\textsf{crs}),\\
  \mathbbm{w}
  \gets
  \mathcal{E}^{\mathcal{A}_1,\mathcal{A}_2}(\textsf{crs},\mathbbm{x})
\end{array}
\right]
\]
}
\end{definition}

% \begin{table*}[t]
% \centering
% \caption{Comparison of our scheme with VeriSimplePIR and SimplePIR.}
% \label{tab:performance_comparison}
% % 如果表格跨栏后依然显得太宽，可以取消下面这行和最后 \end{tabular} 下面那行的注释进行整体缩放
% % \resizebox{\textwidth}{!}{ 
% \begin{tabular}{l | ccc | ccc | ccc}
% \toprule
% \multirow{2}{*}{\textbf{DB Size}} & \multicolumn{3}{c|}{\textbf{Offline Comm. (MiB)}} & \multicolumn{3}{c|}{\textbf{Online Comm. (KB)}} & \multicolumn{3}{c}{\textbf{Online Latency (ms)}} \\
% % \cmidrule(lr) 可以在横线之间留出一点小空隙，视觉效果更精致
% \cmidrule(lr){2-4} \cmidrule(lr){5-7} \cmidrule(l){8-10} 
% & \textbf{SimplePIR} & \textbf{VeriSim.} & \textbf{Ours} & \textbf{SimplePIR} & \textbf{VeriSim.} & \textbf{Ours} & \textbf{SimplePIR} & \textbf{VeriSim.} & \textbf{Ours} \\
% \midrule
% 512 MiB & 209  & 790  & 978  & 209  & 256  & 209  & 68.24   & 92.32   & 69.30   \\
% 1 GiB   & 296  & 1116 & 1337 & 296  & 362  & 296  & 115.83  & 158.72  & 115.90  \\
% 4 GiB   & 591  & 2224 & 2662 & 592  & 724  & 592  & 360.09  & 560.68  & 358.60  \\
% 8 GiB   & 836  & 3149 & 3766 & 836  & 1024 & 836  & 661.09  & 1023.48 & 657.70  \\
% 16 GiB  & 1183 & 4460 & 5323 & 1182 & 1448 & 1182 & 1238.02 & 1810.52 & 1244.10 \\
% \bottomrule
% \end{tabular}
% % }
% \end{table*}

\newpage
\section{Ethics Considerations}
This work does not involve human subjects, personal data, user studies, vulnerable populations, or experimentation on operational systems. The research focuses on the design, analysis, and implementation of cryptographic techniques for verifiable private information retrieval. We are not aware of any specific ethical concerns associated with this work.

\newpage % The Meta-Review should at least start on a new column

\section{Meta-Review}

The following meta-review was prepared by the program committee for the 2027
IEEE Symposium on Security and Privacy (S\&P) as part of the review process as
detailed in the call for papers.

\subsection{Summary}
This paper introduces a new verifiable PIR protocol whose performance roughly matches that of SimplePIR, a highly performant non-verifiable PIR protocol. The key idea is to replace SIS-based commitments with proofs based on error correcting code-based commitments which admit very efficient proofs. 

\subsection{Scientific Contributions}
\begin{itemize}
\item Provides a Valuable Step Forward in an Established Field
%\item text
\end{itemize}

\subsection{Reasons for Acceptance}
\begin{enumerate}
\item The top level performance result is significant, and the evaluation data successfully backs up the claim. 
\item The paper contains both high-level intuition while describing the scheme and detailed, low level arguments for reasoning about its security. 
\item The range proofs included in the paper are of independent interest.
\end{enumerate}

%\subsection{Noteworthy Concerns} % Exclude if your meta-review does not have noteworthy concerns
%\begin{enumerate} % Enumerate environment is not necessary if there is only one
%\item text
%\item text
%\end{enumerate}

%\section{Response to the Meta-Review} % Optional

%Less than 500 words response to the meta-review. The response to the
%meta-review is optional. Provide a response if you disagree with the
%meta-review. Shepherds will only deny responses to meta-reviews if they are too
%long or are abusive / inappropriate.

%\end{appendix}

% that's all folks
\end{document}